\documentclass[english,aps,prl,preprint, showpacs, groupedaddress, amsmath, keywords, showpacs]{revtex4-2}
\usepackage{ulem}
\usepackage{fancyhdr,graphicx}
\usepackage{bm}
\usepackage{color}
\usepackage{amsfonts}
\usepackage{times}
\usepackage{amsmath}
\usepackage{graphicx}
\usepackage{float}
\usepackage{parskip} 
\usepackage{natbib}
\usepackage{chemformula} 
\usepackage[colorlinks=true, citecolor=red]{hyperref}
\setcitestyle{journalcolor= blue}
\usepackage{chemformula} 
\usepackage[T1]{fontenc} 

\usepackage{amsmath} 

\renewcommand{\vec}[1]{\boldsymbol{#1}}

\usepackage{babel}

\begin{document}
	\title{Higher-order Bragg  gaps in the electronic band structure of bilayer graphene renormalized by recursive supermoir\'{e} potential}
	\author{Mohit Kumar Jat$^1$$^\dagger$, Priya Tiwari$^2$$^\dagger$, Robin Bajaj$^1$$^\dagger$, Ishita Shitut$^1$, Shinjan Mandal$^1$, Kenji Watanabe$^3$, Takashi Taniguchi$^4$, H. R. Krishnamurthy$^1$}
	\author{Manish Jain$^1$}
	\email{mjain@iisc.ac.in}
	\author{Aveek Bid$^1$}
	\email{aveek@iisc.ac.in}
	\affiliation{$^1$Department of Physics, Indian Institute of Science, Bangalore 560012, India \\
		$^2$Braun Center for Submicron Research, Department of Condensed Matter Physics, Weizmann Institute of Science, Rehovot, Israel \\
		$^3$ Research Center for Functional Materials, National Institute for Materials Science, 1-1 Namiki, Tsukuba 305-0044, Japan \\
		$^4$ International Center for Materials Nanoarchitectonics, National Institute for Materials Science, 1-1 Namiki, Tsukuba 305-0044, Japan\\
		$^\dagger$ These authors contributed equally.}

	\begin{abstract}
		This letter presents our findings on the recursive band gap engineering of chiral fermions in bilayer graphene doubly aligned with hBN. By utilizing two interfering moir\'{e} potentials, we generate a supermoir\'{e} pattern which renormalizes the electronic bands of the pristine bilayer graphene, resulting in higher-order fractal gaps even at very low energies. These Bragg gaps can be mapped using a unique linear combination of periodic areas within the system. To validate our findings, we used electronic transport measurements to identify the position of these gaps as functions of the carrier density and establish their agreement with the predicted carrier densities and corresponding quantum numbers obtained using the continuum model. Our work provides direct experimental  evidence of the quantization of the area of quasi-Brillouin zones in supermoir\'{e} systems. It fills essential gaps in understanding the band structure engineering of Dirac fermions by a recursive doubly periodic superlattice potential.
	\end{abstract}

	\maketitle

	\section{Introduction}

	Heterostructures of graphene encapsulated between two thin, rotationally misaligned hBN flakes form a stimulating platform for probing topological phases of matter~\cite{gonzalez2021topological,song2015topological,ponomarenko2013cloning,Wang2015,doi:10.1063/1.5094456,Chen2020}. The difference in the lattice constants of hBN and graphene and the angular misalignment between the layers generate two distinct long-wavelength moir\'{e} superlattices at the top and bottom interfaces of graphene with hBN~\cite{Yankowitz2012, Yankowitz2016,doi:10.1126/sciadv.abd3655,Finney2019}. The interference between these patterns forms a supermoir\'{e} structure with multiple complex real-space periodicities, often with a  spatial range larger than that of hBN/graphene moir\'{e} at each interface~\cite{PhysRevB.103.075122,doi:10.1021/acs.nanolett.9b04058, PhysRevB.104.035306,Leconte_2020, PhysRevB.102.045409,doi:10.1126/sciadv.aay8897,doi:10.1126/sciadv.aay8897,sun2021, PhysRevB.103.115419, Yankowitz2019}. The supermoir\'{e} potential (caused by atomic scale modulation of the carbon-carbon hopping amplitudes by the spinor graphene-hBN interaction potential)  effectively folds the graphene band over a smaller Brillouin zone while retaining its original honeycomb structure~\cite{Mayo_2020}. To a first-order, this results in additional, finite-energy split moir\'{e} gaps (SMG) in the graphene dispersion~\cite{doi:10.1021/acs.nanolett.9b04058,Yankowitz2012,Dean2013,doi:10.1126/science.1237240,doi:10.1073/pnas.1424760112, PhysRevB.102.045409,PhysRevB.90.155406, PhysRevB.89.205414, PhysRevB.87.245408,doi:10.1126/science.1237240, Yankowitz2012}. It was recently realized that the superlattice-induced Bragg reflection at the mini Brillouin zone boundaries has additional subtler effects on the electronic dispersion of graphene to arbitrary low energies manifested in the formation of a family of Bragg gaps, van Hove singularities, and even possibly flat bands~\cite{ Leconte_2020,doi:10.1021/acs.nanolett.9b04058, Moriya2020}. Studying these high-order mini-bands in graphene/hBN moir\'{e} superlattice is essential for a detailed understanding of the emergent quantum properties of Dirac fermions in a periodic non-scalar potential~\cite{doi:10.1073/pnas.1424760112, PhysRevB.102.045409}.

	Recent momentum-space low-energy continuum model calculations (valid in the low-energy regime of interest~\cite{PhysRevLett.125.116404,doi:10.1021/acs.nanolett.9b04058,PhysRevB.106.205134}) predict that the positions of these Bragg gaps form a  fractal pattern reminiscent of the Hofstadter butterfly~\cite{PhysRevB.104.035306}. Consequently, the number density of charge carriers at which Bragg scattering (with supermoir\'{e} harmonics) occurs can be described by a unique set of integers called zone quantum numbers~\cite{PhysRevB.104.035306, PhysRevResearch.4.013028}. These are associated with the quasi-Brillouin Zones (qBZ) formed by the multiple reciprocal lattice vectors of the supermoir\'{e} lattice, and their existence implies quantization of the momentum space area of the qBZ~\cite{PhysRevResearch.4.013028}. The zone quantum numbers are topological invariants of the system intimately related to the second Chern numbers~\cite{PhysRevB.104.035306, PhysRevB.101.041112}.  Despite concrete theoretical studies, this very important prediction of the existence of topological invariants in a quasiperiodic crystal like a moir\'{e} superlattice remains experimentally unverified.

	In this Letter, we provide direct experimental evidence of the quantization of the momentum-space area of qBZ in a supermoir\'{e} system. We achieve this through transport measurements and theoretical calculations in doubly aligned high-mobility heterostructures of bilayer graphene (BLG) with hBN. From combined measurements of quantum oscillations, longitudinal resistance $\mathrm{R_{xx}}$ and transverse resistance $\mathrm{R_{xy}}$, we observe and identify a multitude of higher-order Bragg gaps of the supermoir\'{e} structure; these had escaped detection in previous studies~\cite{doi:10.1021/acs.nanolett.8b05061,doi:10.1021/acs.nanolett.9b04058, PhysRevB.104.035306, Leconte_2020, PhysRevB.102.045409,doi:10.1126/sciadv.aay8897,doi:10.1126/sciadv.aay8897,sun2021}. Our continuum-model-based band structure calculations map these gaps uniquely to the  zone quantum numbers of the underlying supermoir\'{e} lattice. Additionally, our analysis traces the origin of several unexplained experimental features in similar graphene/hBN supermoir\'{e} systems reported in recent publications~\cite{doi:10.1126/sciadv.aay8897} (see Supplementary Material S7).


	\subsection{Device characteristics}
	\begin{figure*}[t]
		\includegraphics[width=\textwidth]{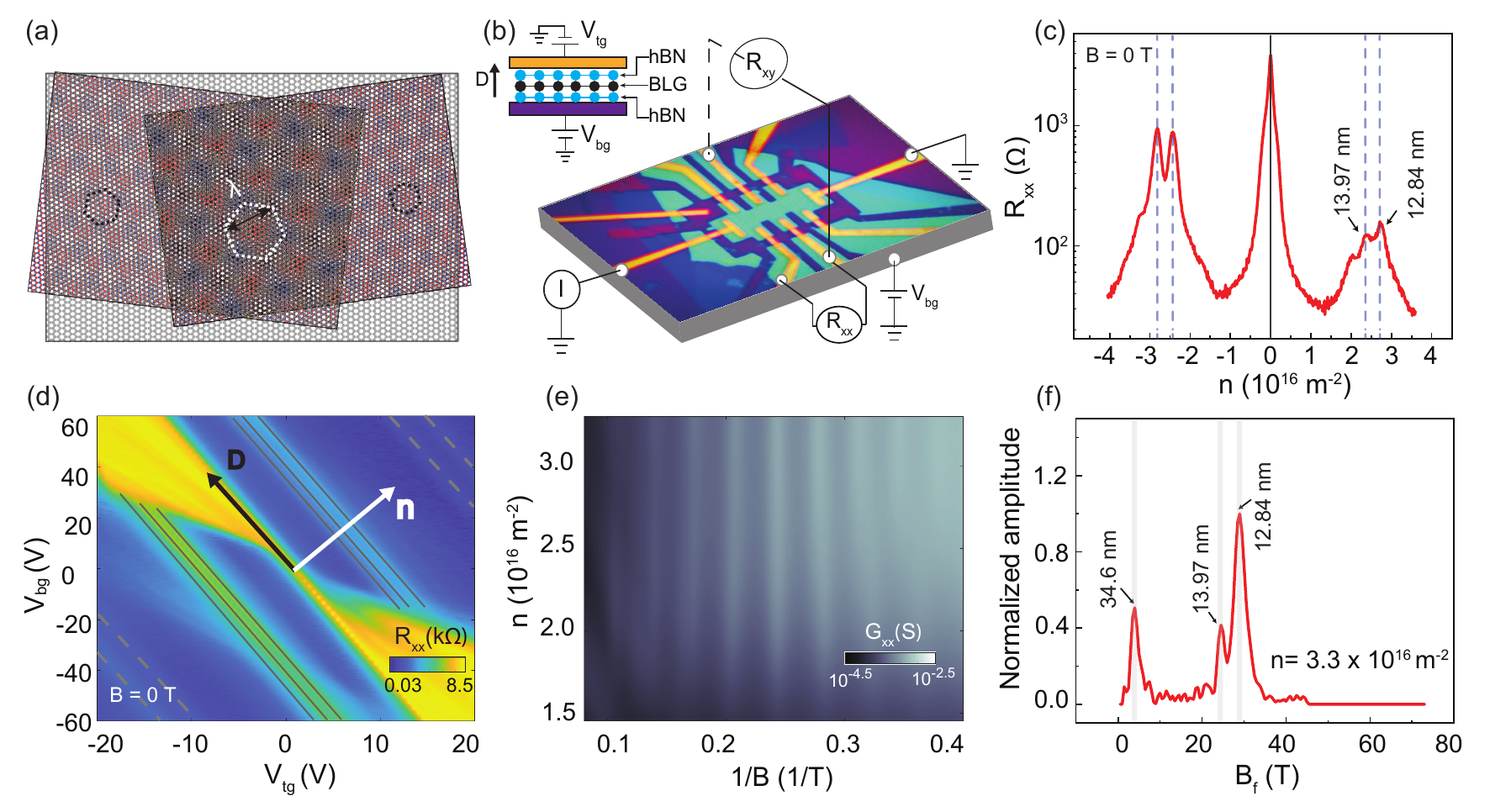}
		\small{\caption{ \textbf{Characteristics of the double moir\'{e} device.} (a) Schematic of doubly aligned BLG with top and bottom hBN. The black and the white hexagons mark the primary moir\'{e} and supermoir\'{e} plaquettes, respectively. (b) An optical image of the device (before adding the top gate) labeled with the measurement configuration. Top inset: Schematic of the layer-stacking, with the direction of increasing displacement field $\mathrm{D}$ marked. (c) Plot of the longitudinal resistance $\mathrm{R_{xx}(B=0)}$ as a function of  $n$; the solid black line marks the position of the charge neutrality point, and the dashed blue lines mark the secondary Dirac points. (d) Map of $\mathrm{R_{xx}}$ as a function of the back gate, $\mathrm{V_{bg}}$ and top gate voltage $\mathrm{V_{tg}}$. (e) Map of the Brown Zak oscillations measured in the $n-1/B$ plane. (f) Fourier spectrum of the BZ oscillation for $n=3.3\times 10^{16}~\mathrm{m^{-2}}$, peaks are marked with the corresponding moire super-lattice wavelengths. }

			\label{fig:fig1}}
	\end{figure*}

	Heterostructures of BLG doubly aligned with hBN with twist angles less than $0.5^\circ$ were fabricated using the dry transfer technique~\cite{pizzocchero2016hot,wang2013one} (see Supplementary Materials S1). The device is in a dual-gated field-effect transistor architecture, allowing independent control on the charge carrier density $n$ and displacement field ${D}$. A contour plot of $R_{xx}$ (Fig.~\ref{fig:fig1}(d)) shows, in addition to the non-dispersive pair of secondary Dirac points, a pronounced asymmetry in the data in the $n-D$ plane. The data matches recent observations in BLG/hBN double moir\'{e} superstructures, where the region with asymmetric conductance was shown to be a  ferroelectric phase~\cite{Zheng2020,Niu2022}; we leave a detailed study of this region of the phase space to the future.

	The appearance of split SMG at $n_b =\pm 2.36\times 10^{16}~\mathrm{m^{-2}}$ and $n_t =\pm 2.80\times 10^{16}~\mathrm{m^{-2}}$ indicates the alignment of the BLG with both the bottom and top hBN layers (Fig.~\ref{fig:fig1}(c-d)). This conclusion is reinforced by the presence of multiple frequencies (at $B_f =$ 24.5~T, 29~T, and 4~T) in the Brown Zak oscillations measured at 100~K (Fig.~\ref{fig:fig1}(e)).  These oscillations, originating from the recurring Bloch states in the superlattice, manifest in the map of magnetoconductance in the $n - 1/B$ plane as dark streaks with positions independent of $n$ and  periodicity $f$  related to the real-space area $S$ of the superlattice by $f =S/\phi_0$, where $\phi_0 = h/e$ is the flux quantum\cite{PhysRevB.14.2239,doi:10.1126/science.aal3357, Barrier2020,Huber2022}. The carrier densities (considering two-fold spin and two-fold valley degeneracies) that fill the two first-order moir\'{e} sub-bands are calculated from  $f = 24.5$~T and 29~T to be $2.36\times 10^{16}~\mathrm{m^{-2}}$ and $ 2.8\times 10^{16}~\mathrm{m^{-2}}$. These numbers match exactly with $n_b$ and $n_t$ identifying these two  Brown Zak oscillation frequencies to be associated with the moir\'{e} formed with hBN at the bottom and top interfaces of BLG, respectively. The corresponding moir\'{e} wavelengths are $\lambda_b = 13.97~\mathrm{nm}$ and $\lambda_t = 12.84~\mathrm{nm}$, respectively.

	We ascertain that both the top- and bottom-hBN crystals have the same relative rotation direction with the intervening graphene layer, with twist angles $\theta_b = 0.03^\circ$ (between bottom hBN and graphene) and $\theta_t = 0.44^\circ$ (between top hBN and graphene) (see Supplementary Materials S2). The very small values of the twist angles place our device in the commensurate limit~\cite{woods2014commensurate}.  We note that the Brown Zak frequency $f_s$ yields $n_s = 0.39 \times 10^{16}~\mathrm{m^{-2}}$ -- this number density corresponds to a real-space wavelength of $35~\mathrm{nm}$ which is the size of the supermoir\'{e} unit cell in our heterostructure.

	\subsection{Continuum Hamiltonian}
	\begin{figure*}[t]
		\begin{center}
			\includegraphics[width=1\textwidth]{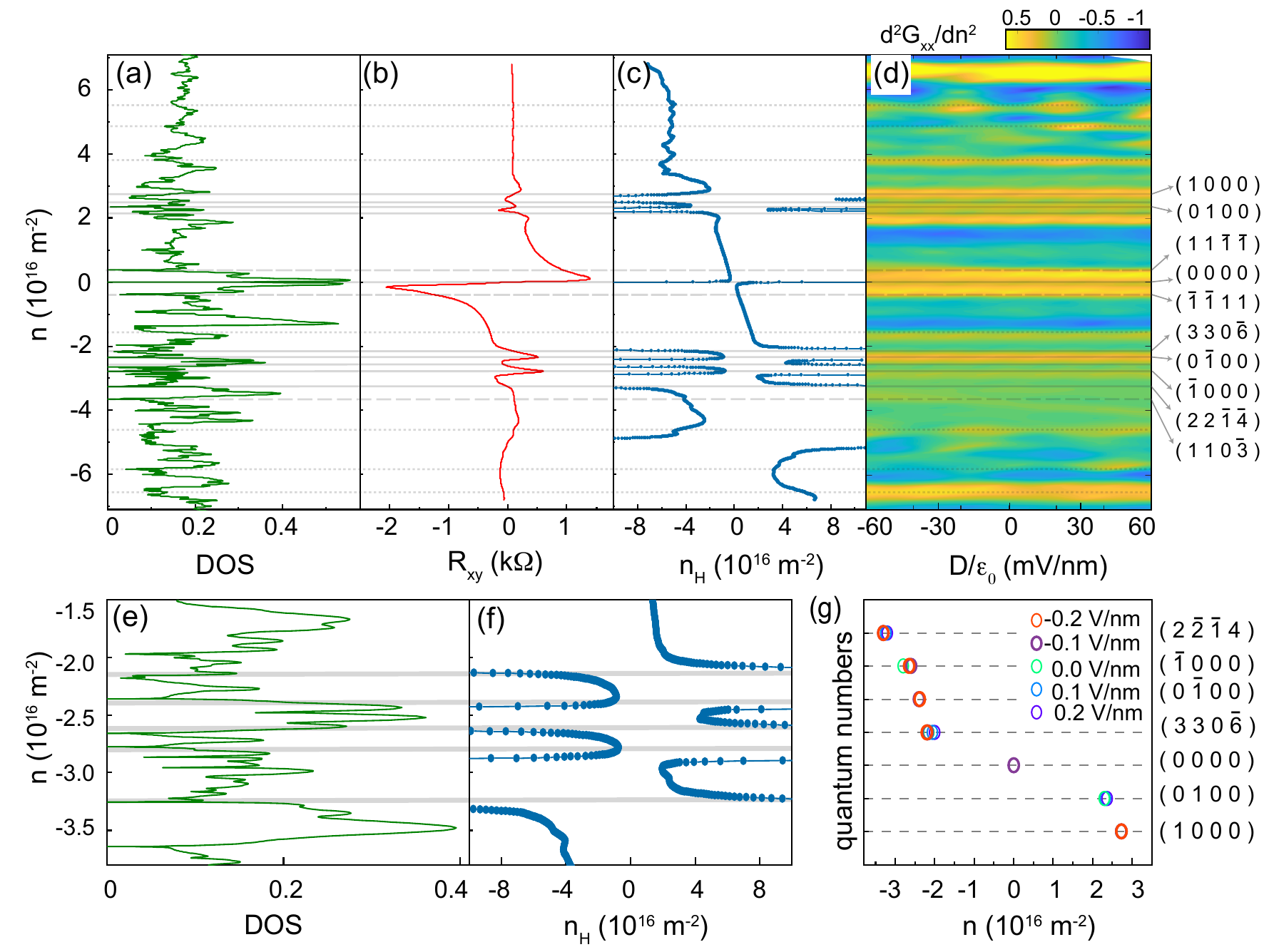}
			\small{\caption{\textbf{Experimentally obtained and theoretically calculated Bragg gaps.}  (a) Plot of the calculated density of states (DOS) for $\theta^{b}= 0.026^{\circ}$ and $\theta^{t}=0.44^{\circ}$. Horizontal lines represent prominent dips in the DOS.  (b) Plot of transverse resistance $\mathrm{R_{xy}}$ versus $n$ measured at $B = 0.7$~T and $T=2$~K. (c) Plot of hall carrier density $n_{H}$ versus $n$. (d) Map of the normalized  $d^2\mathrm{G_{xx}(B=0)}/dn^2$ in the $n-\mathrm{D}$ plane; the data have been plotted on a logarithmic scale. The indices of the Bragg gaps expected at these values of $n$ are marked on the right. (e-f) Zoomed-in plots of DOS and $\mathrm{n_H}$ versus $n$ in a narrow range on the hole-side. (g) Plot of the position of the Bragg numbers versus $n$ over a range of displacement fields.}
				\label{fig:fig2}}
		\end{center}
	\end{figure*}

	\begin{figure*}[t]
		\begin{center}
			\includegraphics[width=1\textwidth]{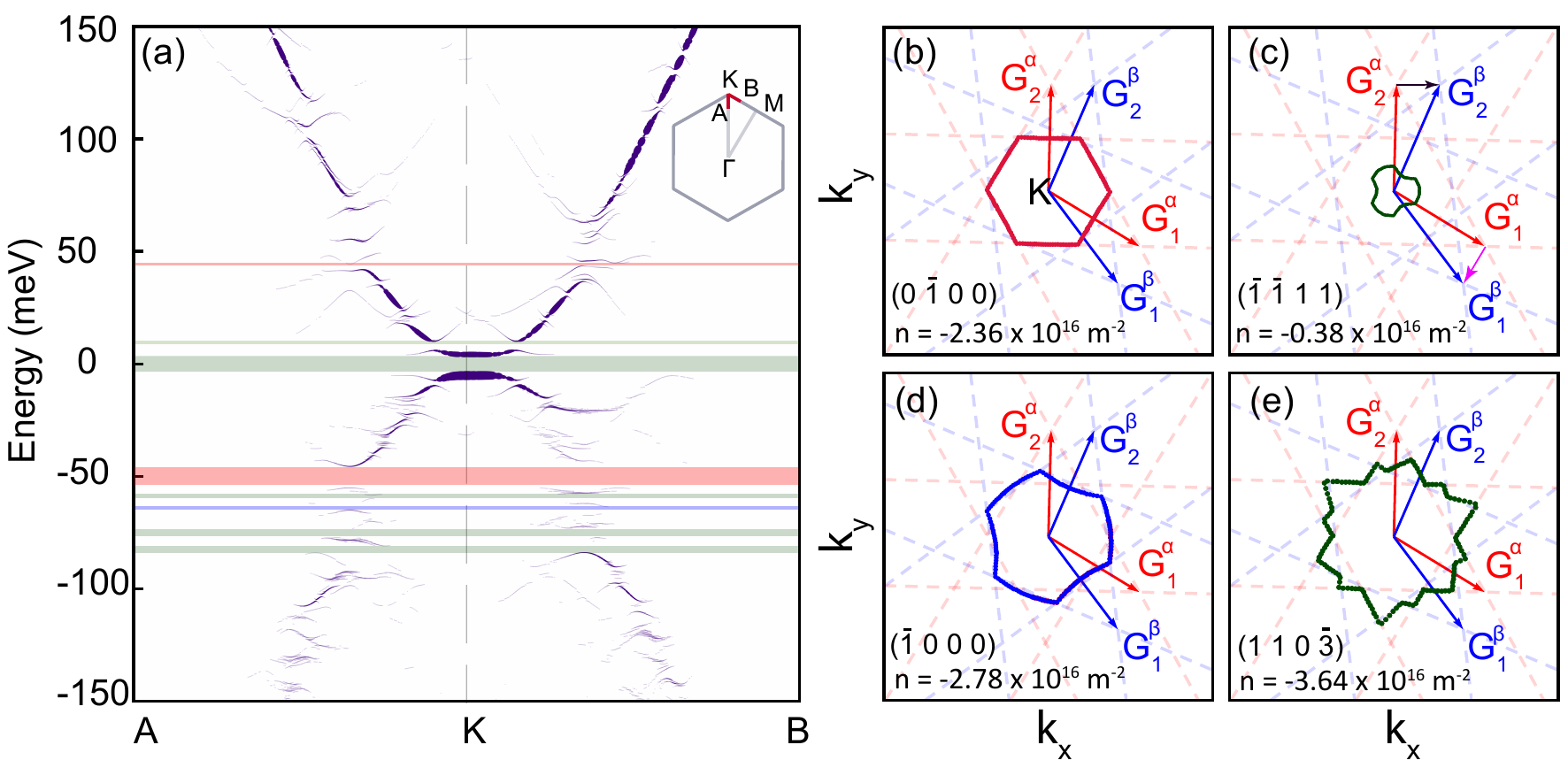}
			\small{\caption{\textbf{Unfolded band structure and qBZs} (a) Unfolded band structure along the path $\mathrm{AK B}$ is shown in the subplot. Primary gaps from the top and bottom moir\'{e} are shown in red and blue. Gaps arising from both layers, \textit{i.e.} supermoir\'{e} gaps, are shown in green. (b-e) Plots of the calculated qBZ for the Bragg gaps at number densities $ -2.36\times 10^{16} \mathrm{m^{-2}}$,  $-0.38\times 10^{16} \mathrm{m^{-2}}$,  $-2.78\times 10^{16} \mathrm{m^{-2}}$, and  $-3.64\times 10^{16} \mathrm{m^{-2}}$, respectively. The zone quantum numbers are indicated in the insets of each panel. The k-points where the gap opens are shown as dots. The $\mathrm{k_x}$ and $\mathrm{k_y}$ range between [-0.8,0.8] nm$^{-1}$. }
				\label{fig:fig3}}
		\end{center}
	\end{figure*}

	Having established the presence of the supermoir\'{e} structure, we move on to discuss its effect on the bilayer graphene band structure using the Bistritzer MacDonald continuum model~\cite{doi:10.1073/pnas.1108174108} The 4$\times$4 effective Hamiltonian (eliminating the sub lattice basis of hBN using second-order perturbation theory) is written as:
	\begin{eqnarray}
		H_{eff} =
		\begin{bmatrix}
			H_{G}+V_{hBN}^{b} & U_{BLG}^{\dagger} \\
			U_{BLG}                & H_{G}+V_{hBN}^{t}
		\end{bmatrix}
		\label{Eqn:1}
	\end{eqnarray}
	where in the low-energy limit,
	\begin{eqnarray}
		V_{hBN}^{\ell} = U^{\ell\dagger}(-H_{hBN})^{-1}U^{\ell}
		= v_{0} + v_{1}e^{i\xi\vec{G}_{1}^{\ell}.\vec{r}}+
		v_{2}e^{i\xi\vec{G}_{2}^{\ell}.\vec{r}}+v_{3}e^{i\xi\vec{G}_{3}^{\ell}.\vec{r}}
	\end{eqnarray}
	Here $\ell= t,b$ and $\xi = \pm 1$ is the  valley index.  $\vec{G}_{1}^{\ell}$ and $\vec{G}_{2}^{\ell}$ are the reciprocal lattice vectors of the $\ell$ moire and $\vec{G}_{3}^{\ell} = -\vec{G}_{1}^{\ell}-\vec{G}_{2}^{\ell}$ (see top left inset of Fig.~\ref{fig:fig3}). $U_{BLG}$ is the inter-layer potential between the layers of the BLG.

	Fig.~\ref{fig:fig2}(a) shows the theoretically constructed density of states (DOS) versus carrier density plot; the zeros in the DOS correspond to the gaps in the energy spectrum. To gain a physical understanding of the origin of these gaps, we follow the procedure laid out in Ref ~\cite{PhysRevB.104.035306}. Recall that a nearly commensurate system with dual periodicity is defined by a set of four distinct reciprocal wave vectors:  $\vec{G}_{1,2}^{t}$ being the two primitive reciprocal lattice vectors of the moir\'{e} lattice at the top hBN-graphene interface and $\vec{G}_{1,2}^{b}$ those for the second moir\'{e} lattice at the bottom graphene-hBN interface. One can form quasi-Brillouin zones bounded by multiple Bragg planes defined by a linear combination of these four primary reciprocal vectors. The $(m_1,m_2.m_3,m_4)^\mathrm{{th}}$ -- order Bragg-gap appears in the electronic spectrum when the total charge carrier density equals~\cite{doi:10.1021/acs.nanolett.8b05061, PhysRevB.104.035306}:
	\begin{equation}\label{Eqn:bragg-gap}
		n(m_1,m_2,m_3,m_4) = 4\sum_{i=1}^{4} m_i A_i/(2\pi)^2.
	\end{equation}
	Here $A_1 = |\vec{G}_{1}^{b}\times \vec{G}_{2}^{b}|$, $A_2 = |\vec{G}_{1}^{t}\times \vec{G}_{2}^{t}|$, $A_3 = |\vec{G}_{1}^{b}\times \vec{G}_{2}^{t}|$, and $A_4 = |\vec{G}_{1}^{t}\times \vec{G}_{2}^{b}|$ are the areas of the projections of the parallelograms formed by the four reciprocal lattice vectors $\vec{G_i}$. The quantity $\sum_{i,j=1}^{4} m_i A_i$ is the area (in reciprocal space) of the multifaceted quasi-Brillouin zone, and the factor of four on the right-hand side of Eqn.~\ref{Eqn:bragg-gap} arises from the spin and valley degeneracies. The areas $A_i$ for the experimentally obtained twist angle $\theta_b = 0.03^\circ$ and $\theta_t = 0.44^\circ$  are $\mathrm{0.277, \ 0.233, \ 0.181}$ and $\mathrm{0.290}$ nm$^{-2}$.  The integers $m_i$ are zone quantum numbers of the gap and are topological invariants of the system~\cite{PhysRevResearch.4.013028, PhysRevB.104.035306}. Note that this formalism is mathematically identical to that utilized by previous workers based on differences between the multiples of the aligned and rotated reciprocal vectors~\cite{doi:10.1126/sciadv.aay8897,doi:10.1021/acs.nanolett.8b05061, Leconte_2020,doi:10.1021/acs.nanolett.9b04058} with the added advantage of being intuitively transparent.

	\subsection{Experimental observation of the Bragg gaps}

	Using the above formalism, the band gaps corresponding to the densities $n_b$, $n_t$, and $n_s$ are identified to be Bragg gaps with zone quantum numbers  $(0,1,0,0)$, $(1,0,0,0)$, and $(1,1,\bar{1},\bar{1})$ respectively (see Supplementary Materials S4).
	We experimentally obtain the positions of additional Bragg gaps from the transverse resistance ${R_{xy}(B)}$ and the extracted Hall carrier density ${n_H(B)}$, (Fig.~\ref{fig:fig2}(b-c)) measured in the presence of a small, non-quantizing magnetic field $B=0.7$~T. Recall that for small $B$, ${n_H}$ has an apparent divergence at a band gap, and its sign reflects the local band curvature. Thus, for instance, with $E_F>0$, one can have both positive and negative ${n_H}$; a positive (negative) value of ${n_H}$ implies a hole-like (an electron-like) band. The ${n_H(B)}$ measured in our supermoir\'{e} device shows several additional divergences and sign changes, including the expected ones at the CNP ($n=0$) and the two primary moir\'{e} gaps ($\pm n_b$ and $\pm n_t$) (Fig.~\ref{fig:fig2}(b-c)). These additional regions where ${n_H}$ changes sign and diverges are identified to be higher-order Bragg-gaps with specific quantum numbers -- several of them have been indicated in Fig.~\ref{fig:fig2}.

	In addition to these prominent features, the ${n_{H}}$ data in Fig.~\ref{fig:fig2}(b) has several weaker sign changes and inflections. These are identified as additional Bragg gaps. To substantiate this claim, in Fig.~\ref{fig:fig2}(d), we plot a map of the second derivatives of the longitudinal conductance ${G_{xx}(B=0)}$ in the $n-D$ plane. Every zero (and several prominent non-zero dips) in the calculated DOS (Fig.~\ref{fig:fig2}(a))  is reflected in the experimental data as a  discontinuity in the ${n_H}-n$ plot (Fig.~\ref{fig:fig2}(c)) and as a local maximum in $d^2\mathrm{G_{xx}}/dn^2$ (minima in $\mathrm{G_{xx}}$) (Fig.~\ref{fig:fig2}(d)). Each such gap can be uniquely mapped to Bragg gaps with integer zone quantum numbers (see Supplementary Materials S4) -- we provide specific examples supporting this claim in Fig.~\ref{fig:fig2}(e-f). The fact that the data from measurements of three independent physical quantities (quantum oscillations, Hall resistance, and zero-magnetic-field longitudinal resistance) and from continuum-model-based calculations match emphasizes the validity of our analysis. We note in passing that the positions of these gaps in number density are independent of applied small displacement fields  (Fig.~\ref{fig:fig2}(g)).

	From the activated temperature-dependent resistance data, we extract the band-gap at CNP to be $6~\mathrm{meV}$ at zero displacement field. This value is in the same range as our
	theoretically calculated band gap $8.6 ~\mathrm{meV}$ and is in agreement with the recent theoretical work in supermoir\'{e} system ~\cite{PhysRevB.106.205134} and experimental studies in transport~\cite{PhysRevB.103.115419}. The energy gaps at  the primary moir\'{e} gaps are extracted to be $E_b = 1.46~\mathrm{meV}$ and $E_t = 3.39 ~\mathrm{meV}$ respectively.

	\subsection{Quasi Brillouin Zones}

	The electronic carrier densities at which we observe the gaps in our doubly-periodic 2D system are related to the areas of the underlying qBZ. In order to identify these zone boundaries, we find the $k$ points at which the gaps open. One can observe the gap opening points by unfolding the supermoir\'{e} band structure to the unit cell of the BLG. We modulate the strength of top and bottom moir\'{e} potential in the reduced Hamiltonian (Eqn.~\ref{Eqn:1}) with strength parameter $\eta$ ranging from 0 to 1 (See Supplementary Materials S5 to identify the gap-opening points accurately). The unfolded band structure (Fig.~\ref{fig:fig3}(a)) can be seen along a given $k$-path using unfolded spectral weights as:
	\begin{equation}
		A(\vec{q},\epsilon) = \sum_{n\vec{k}}\sum_{X}\left|\langle\vec{q},X|\psi_{n\vec{k}}\rangle \right|^{2}\delta(\epsilon-\epsilon_{n\vec{k}})
	\end{equation}
	where $X = A_{1}, B_{1}, A_{2}, B_{2}$ denote the atoms of bilayer graphene, $|\psi_{n\bf{k}}\rangle$ and $\epsilon_{n\bf{k}}$ denote the eigenstates and eigenvalues, respectively,  $\vec{q}$ is the crystal momentum in the bilayer graphene unit cell BZ. The $\vec{q}$ is related to $\vec{k}$ in the supermoir\'{e} BZ with a moir\'{e} reciprocal lattice vector $\vec{G_{SM}}$ via the relation $\vec{q} = \vec{k}+\vec{G_{SM}}$~\cite{Mayo_2020}.

	Figs.~\ref{fig:fig3}(b-e) shows the calculated qBZ for a few zone quantum numbers using the above procedure. These shapes and the corresponding zone quantum numbers have simple geometrical interpretations. Consider, for example, the qBZ of the supermoir\'{e} cell plotted in Fig.~\ref{fig:fig3}(c); it is formed by the reciprocal lattice vectors $\vec{G_1^b}-\vec{G_1^t}$ and $\vec{G_2^b}-\vec{G_2^t}$. The area of this qBZ can be expressed as:
	\begin{equation}
		\begin{split}
			(\vec{G_1^b}-\vec{G_1^t}) \times (\vec{G_2^b}-\vec{G_2^t}) & = (\vec{G_1^b}\times \vec{G_2^b})+(\vec{G_1^t}\times \vec{G_2^t})-(\vec{G_1^b}\times \vec{G_2^t})-(\vec{G_1^t}\times \vec{G_2^b})
			\\
			& = |\bf{A_1}|+|\bf{A_2}|-|\bf{A_3}|-|\bf{A_4}|
		\end{split}
	\end{equation}
	This gives the zone quantum numbers of the qBZ of the supermoir\'{e} to be $(1,1,\bar{1},\bar{1})$ (see Eqn.~\ref{Eqn:bragg-gap}) with the number density required to fill the band $n_s =0.39\times 10^{16}~\mathrm{m^{-2}}$.
	We thus find the area of the supermoir\'{e} qBZ  arrived at using two very  different theoretical routes (continuum model calculations and band geometric considerations) to be in excellent agreement with that extracted from measured Brown Zak oscillations.

	A closer inspection reveals that several of the qBZ are distorted hexagons; two examples are provided in Fig.~\ref{fig:fig3}(c-d). The source of this distortion can be traced back to the triangular symmetry of the constant energy contours of bilayer graphene energy dispersion (See Supplementary Materials S6). Fig.~\ref{fig:fig3}(e) shows an example of the fractal or flower-like qBZ for higher-order gap Bragg  predicted for doubly aligned graphene~\cite{PhysRevB.104.035306}.

	We note in passing that throughout the above discussion, we have avoided any mention of the strength of the interlayer coupling. As noted in previous studies, the interlayer coupling strength affects only the magnitude of the Bragg gaps, leaving their positions unaffected~\cite{doi:10.1021/acs.nanolett.9b04058}.

	To summarize, we show that the low-energy dispersion of bilayer graphene can be significantly altered by the supermoir\'{e} potential. Our transport measurements combined with theoretical analysis provide an elegant physical picture of the Bragg gaps opening in the moir\'{e} spectrum and help identify several higher-order Bragg gaps with well-defined zone quantum numbers. Our experimental results match extremely well with the predictions of the subtle effects of nearly commensurate supermoir\'{e} structures on graphene bands. Interestingly, our calculations establish that the qBZ of the supermoir\'{e} lattice in bilayer graphene is $\mathcal{C}_3$ symmetric, making it an ideal system to host intrinsic Berry curvature dipoles. Further experiments and theoretical calculations that include interaction effects are required to understand the entire physics of this fascinating material and realize its full potential.

	\section{Methods}
	\subsection{Device fabrication and measurement}
	Devices of bilayer graphene (BLG) heterostructures doubly aligned with single crystalline hBN were fabricated using a dry transfer technique (for details, see Supplementary materials S1). Raman spectroscopy and AFM were used to determine the number of layers and thickness uniformity, respectively. The heterostructure was aligned to form a moiré superstructure with less than $1^\circ$ misalignment. The devices were patterned using electron beam lithography, followed by reactive ion etching and thermal deposition of Cr/Au contacts. The dual-gated device architecture allows for independent tuning of charge carrier density and displacement field. The capacitance values of the top gate and back gate were extracted from quantum hall measurements. Measurements were done in a Cryogen-free refrigerator (with a base temperature of $\mathrm{2 K}$ and magnetic field up to $\mathrm{14 T}$) at low frequency using standard low-frequency measurement techniques.

	\section{Data availability}
	The authors declare that the data supporting the findings of this study are available within the main text and its Supplementary Information. Other relevant data are available from the corresponding author upon reasonable request.

	\section{Acknowledgment}
	A.B. acknowledges funding from U.S. Army DEVCOM Indo-Pacific (Project number: FA5209   22P0166) and Department of Science and Technology, Govt of India (DST/SJF/PSA-01/
	2016-17). M.J. and H.R.K. acknowledge the
	National Supercomputing Mission of the Department of
	Science and Technology, India, and the Science and Engineering Research Board of the Department of Science
	and Technology, India, for financial support under Grants
	No. DST/NSM/R$\&$D\_HPC Applications/2021/23 and
	No. SB/DF/005/2017, respectively. M.K.J. and R.B. acknowledge the funding from the Prime minister's research fellowship (PMRF), MHRD.

	\section{Author contributions}
	M.K.J., P.T., and A.B. conceived the idea of the study,  conducted the measurements, and analyzed the results. T.T. and K.W. provided the hBN crystals. M.J., R.B., S.M., I.S., and H.R.K. developed the theoretical model. All the authors contributed to preparing the manuscript.

	\section{Competing interests}
	The authors declare no competing interests.

	\clearpage

	\clearpage

	\section*{Supplementary Materials}

	\renewcommand{\theequation}{S\arabic{equation}}
	\renewcommand{\thesection}{S\arabic{section}}
	\renewcommand{\thefigure}{S\arabic{figure}}
	\renewcommand{\thetable}{S\arabic{table}}
	\setcounter{table}{0}
	\setcounter{figure}{0}
	\setcounter{equation}{0}
	\setcounter{section}{0}

	\section{S1.~\hspace{0.2cm} Device fabrication}
\begin{figure}[H]
		\begin{center}
			\includegraphics[width=0.45\columnwidth]{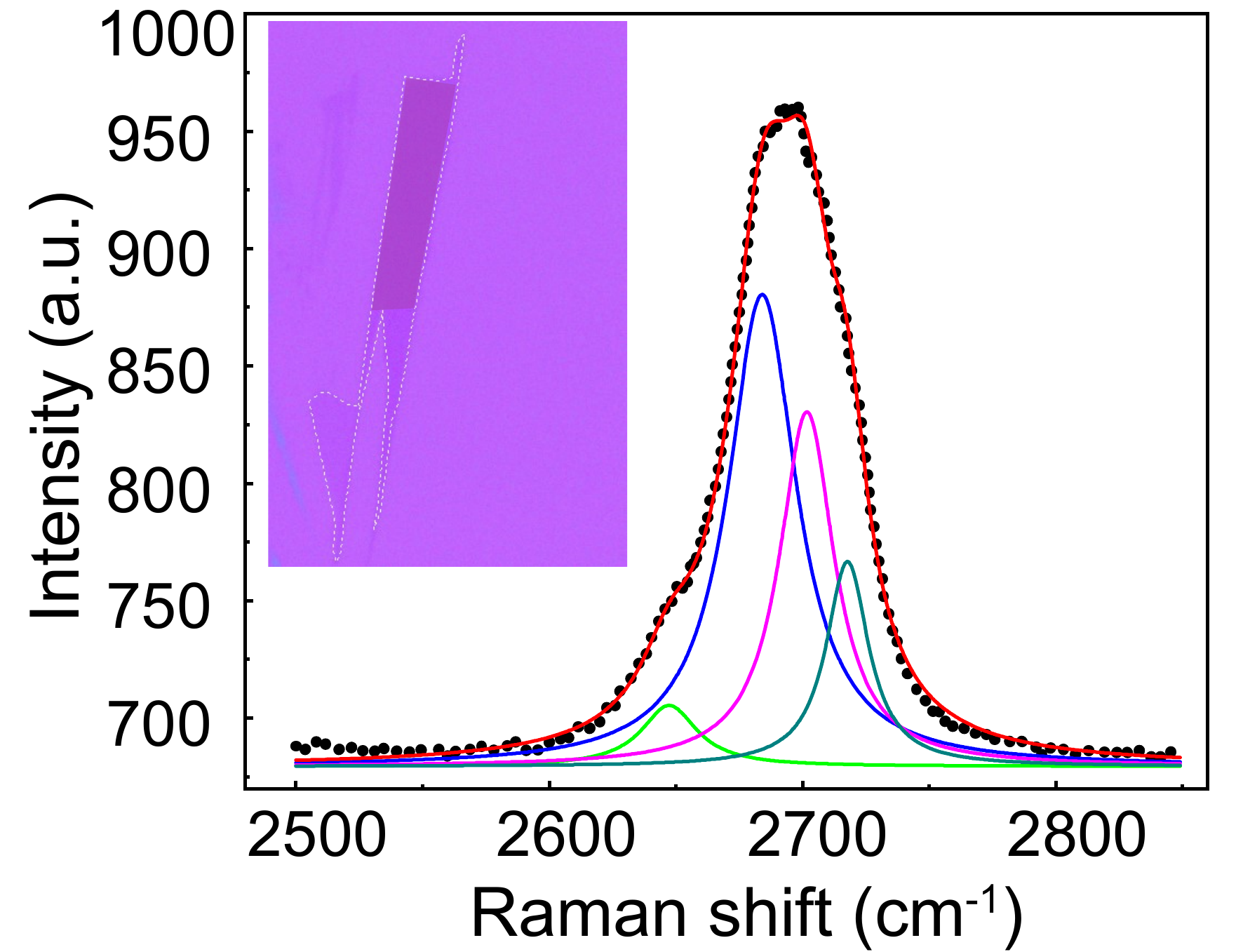}
			\small{\caption{\textbf{Raman spectra}. A plot of the $\mathrm{2D}$ Raman spectra of the bilayer graphene. The black-filled circles are the experimentally measured Raman data. The red solid line is cumulative of the four Lorentzian fitted to it; the four Lorentzian are also individually shown. The inset shows an optical image of the  BLG flake; the shaded region is the part of the BLG used for device fabrication.}
				\label{fig:Raman}}
		\end{center}

	\end{figure}

	We fabricated bilayer graphene (BLG) heterostructures doubly aligned with the single crystalline hBN of thickness 20-25$~\mathrm{nm}$. The flakes were mechanically exfoliated on \ch{Si}/\ch{SiO2} wafers. The number of layers in the graphene flake was determined from Raman spectroscopy (Fig.~\ref{fig:Raman}).  AFM was used to measure the thicknesses of the hBN flakes and ensure their uniformity.

\begin{figure}[t]
		\begin{center}
			\includegraphics[width=0.9\columnwidth]{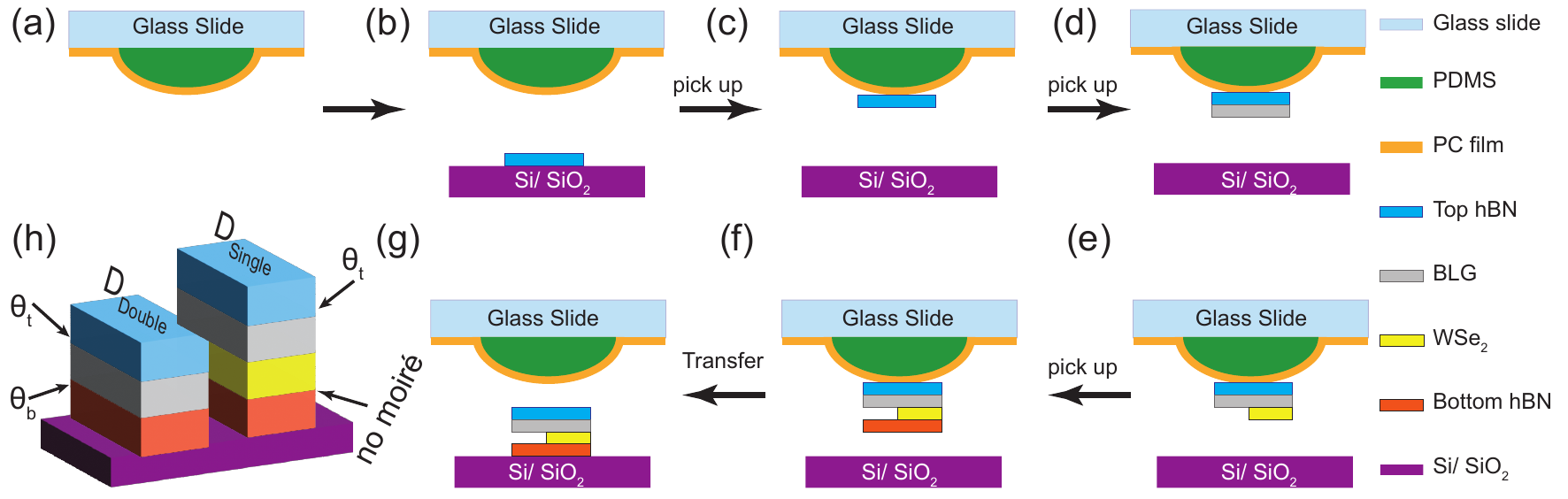}
			\small{\caption{\textbf{Device fabrication schematic} (a) PDMS dome covered with a thin film of PC. (b) Desired top hBN flake is brought in contact with the PC at $120^{\circ}$C. (c) After successful pickup, Si/SiO$_{2}$ wafer is brought down. (d) A BLG flake is picked up.   (e) A WSe$_{2}$ flake is picked up on one-half of the BLG. (f) The bottom hBN flake is picked. (g) The heterostructure is transferred on clean Si/SiO$_{2}$ substrate by melting PC at $190^{\circ}$C. (h) The heterostructure is finally etched in two halves.}
				\label{fig:Devicefab}}
		\end{center}

	\end{figure}

	The constituent layers of the heterostructure were sequentially transferred on each other using a dry transfer technique (Fig.~\ref{fig:Devicefab}). During the transfer, the edges of the crystals were carefully aligned (aiming for less than $1^{\circ}$ misalignment)  to form the moir\'{e} heterostructure.  In the first step, a BLG flake was aligned to an hBN flake at a near-zero angle such that a moir\'{e} superstructure forms between the entire top surface of BLG and the hBN (Fig.~\ref{fig:Devicefab}(d)). Half the BLG flake was then covered with a single-layer \ch{WSe2} (Fig.~\ref{fig:Devicefab}(e)). An hBN flake was then picked up by the stack, ensuring that it had a near-zero angular mismatch with the BLG (Fig.~\ref{fig:Devicefab}(f)). The two halves of the BLG were then separated into two different devices, labelled $\mathrm{D_{double}}$ and $\mathrm{D_{single}}$, respectively (Fig.~\ref{fig:Devicefab}(h)). The BLG in device $\mathrm{D_{double}}$ formed a double moir\'{e} heterostructure with the top- and the bottom-hBN flakes. Device $\mathrm{D_{single}}$, on the other hand, formed a moir\'{e} heterostructure with only the top-hBN. The results presented in the main manuscript are from device $\mathrm{D_{double}}$. The device $\mathrm{D_{single}}$ (which is fabricated from the same BLG and top hBN flake as device $\mathrm{D_{double}}$) forms the ideal control device for comparison of the characteristics of single- and double-moir\'{e} heterostructures of hBN and BLG.

	Electron beam lithography was used to pattern electrical contacts, followed by reactive ion etching (using a mixture CHF$_3$ (40 sscm) + O$_2$ (4 sscm)) and thermal deposition  of Cr/Au (5nm/55nm) to create one-dimensional edge contacts~\cite{wang2013one}. The devices were then etched to define the Hall bar geometry. In the last step, top gates were fabricated using electron beam lithography and thermal deposition of Cr/Au (5nm/55nm). The presence of dual-gated device architecture provides the freedom  to tune the charge carrier density $n$ and  displacement field ${D}$ independently via the relations $n \equiv  (\mathrm{C}_{tg}\mathrm{V}_{tg}+\mathrm{C}_{bg}\mathrm{V}_{bg})/e-n_0$ and ${D} \equiv(\mathrm{C}_{bg}\mathrm{V}_{bg}-\mathrm{C}_{tg}\mathrm{V}_{tg})/2-{D}_0$ (the effective electric field in the system is $D/{\epsilon_{0}}$). Here $n_0$ is the residual charge density due to doping, and ${D}_0$ is the net internal displacement field. $\mathrm{C}_{tg}$ and $\mathrm{C}_{bg}$ are the top and bottom gate capacitance respectively; their values are extracted from quantum hall measurements.

	\section{S2.~\hspace{0.2cm} Determining relative twist between BLG and hBN}

	\subsection{A.~\hspace{0.2cm} Device ${\mathbf{D_{single}}}$ - single moir\'{e} device}
\begin{figure*}
		\includegraphics[width=1\columnwidth]{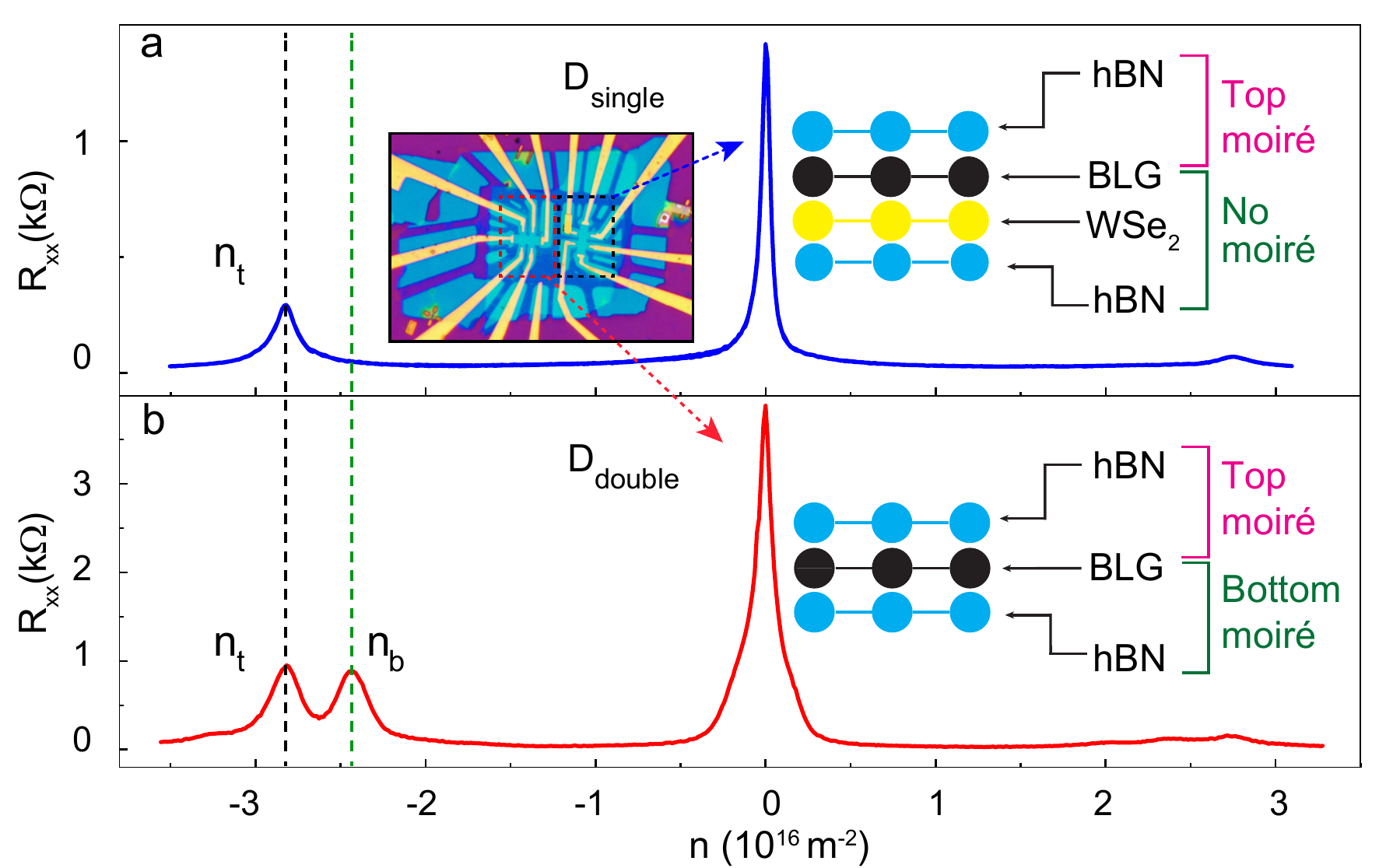}
		\small{\caption{ \textbf{Longitudinal resistance comparison for single and supermoir\'{e}}.(a) Plot of longitudinal resistance $\mathrm{R_{xx}}$ versus $n$ for the single moir\'{e} device $\mathrm{D_{single}}$. The inset shows the schematic of the layer stacking. The resistance peak at $n_{t} = -2.8\times 10^{16}~\mathrm{m^{-2}}$ (shown by the dashed black line) is from the moir\'{e} reconstruction of electronic bands. (b) Plot of $\mathrm{R_{xx}}$ versus $n$ for  the supermoir\'{e} device $\mathrm{D_{double}}$. The inset shows the schematic of the layer stacking.  Two secondary resistance peaks are seen at $n_{b}= -2.36\times 10^{16}~\mathrm{m^{-2}}$ (green dashed line) and $n_{t} = -2.8\times 10^{16}~\mathrm{m^{-2}}$ (black dashed line).}
			\label{fig:fig9}}
	\end{figure*}

	Fig.~\ref{fig:fig9}(a) shows a plot of the longitudinal resistance $\mathrm{R_{xx}}$ versus carrier density $n$ measured for the single-moir\'{e} device $\mathrm{D_{single}}$. The resistance peak at $n{_t} = \pm 2.8\times 10^{16}~\mathrm{m^{-2}}$ is a  consequence of the opening of a moir\'{e} gap (MG) at this number density. The moir\'{e} wavelength was calculated using the relation:
	\begin{eqnarray}
		\lambda^2 = (8/\sqrt{3}) n \label{Eqn:moirelength}
	\end{eqnarray}
	to be $\lambda{_t} = 12.84~\mathrm{nm}$. The twist angle between the BLG and the top hBN was estimated using the general relation between twist angle and the moir\'{e} wavelength~\cite{PhysRevB.90.155406,doi:10.1126/science.1237240, doi:10.1021/acs.nanolett.8b05061}:
	\begin{eqnarray}
		\mathrm{\lambda = \frac{(1+\epsilon)a}{[\epsilon^2 +2(1+\epsilon)(1-cos(\theta))]^{1/2}}}\label{Eqn:moireangle}
	\end{eqnarray}
	Here $\mathrm{a} = 0.246~\mathrm{nm}$ is the lattice constant of graphene, $\epsilon =0.018$ is the lattice mismatch between the hBN and graphene, and $\theta$ is the relative rotational angle between the two lattices. We find the  magnitude of the twist angle between the BLG and the top hBN to be $|\theta{_t}| = 0.44^\circ$.

	\subsection{B.~\hspace{0.2cm} Device $\mathbf{D_{double}}$ -  supermoir\'{e} device}

	The plot of $\mathrm{R_{xx}}$ versus $n$ for the device $\mathrm{D_{double}}$ shows split moir\'{e} gaps (SMG) at $n_{b}= \pm 2.36\times 10^{16}~\mathrm{m^{-2}}$ and $n_{t} = \pm 2.80\times 10^{16}~\mathrm{m^{-2}}$ (Fig.~\ref{fig:fig9}(b)). This indicates an alignment of the BLG with both the top and bottom hBN layers.  From the positions of the SMG, we extract the two moir\'{e} wavelengths to be $\lambda{_b}=13.96~\mathrm{nm}$ and $\lambda{_t} = 12.84~\mathrm{nm}$ (Eqn.~\ref{Eqn:moirelength}). The magnitudes of the corresponding twist angles are $|\theta{_b}| = 0.03^\circ$ and $|\theta{_t}| = 0.44^\circ$ (Eqn.~\ref{Eqn:moireangle}). Recalling that (1) these two devices have common BLG and top hBN flakes and (2) the top hBN forms a moir\'{e} with twist angle $|\theta{_t}| = 0.44^\circ$ with the BLG (as seen in the previous section from the data for device $\mathrm{D_{single}}$); we conclude that the moir\'{e} formed between the bottom hBN and BLG has a moir\'{e} wavelength $\lambda{_b}=13.96$ nm and a twist angle  $|\theta{_b}| = 0.03^\circ$.

	We find the frequencies of the Brown-Zak oscillations to be $f_b = 24.5$~T ($\lambda{_b} = 13.97~\mathrm{nm}$) and $f_a = 29$~T ($\lambda{_t} = 12.84~\mathrm{nm}$) (see main manuscript). The corresponding carrier densities which satisfy the conditions for BZ oscillations are $|n_{b}|= 2.36\times 10^{16}~\mathrm{m^{-2}}$ and $|n_{t}| = 2.8\times 10^{16}~\mathrm{m^{-2}}$. These values match extremely well with the locations of the secondary moir\'{e} peaks in Fig.~\ref{fig:fig9}(b).

	\subsection{C.~\hspace{0.2cm} Size of the supermoir\'{e} cell}

	The number density at which supermoir\'{e} gap opens is given analytically by Ref.~\cite{doi:10.1126/sciadv.aay8897}:
	\begin{eqnarray}
		\mathrm{n_s = -\frac{16[cos(\theta{_b}-\theta{_t})-1]}{\sqrt3 a^2 (1+\delta)^{2}}}.\label{Eqn:nsm}
	\end{eqnarray}
	\begin{figure}
		\begin{center}
			\includegraphics[width=0.45\columnwidth]{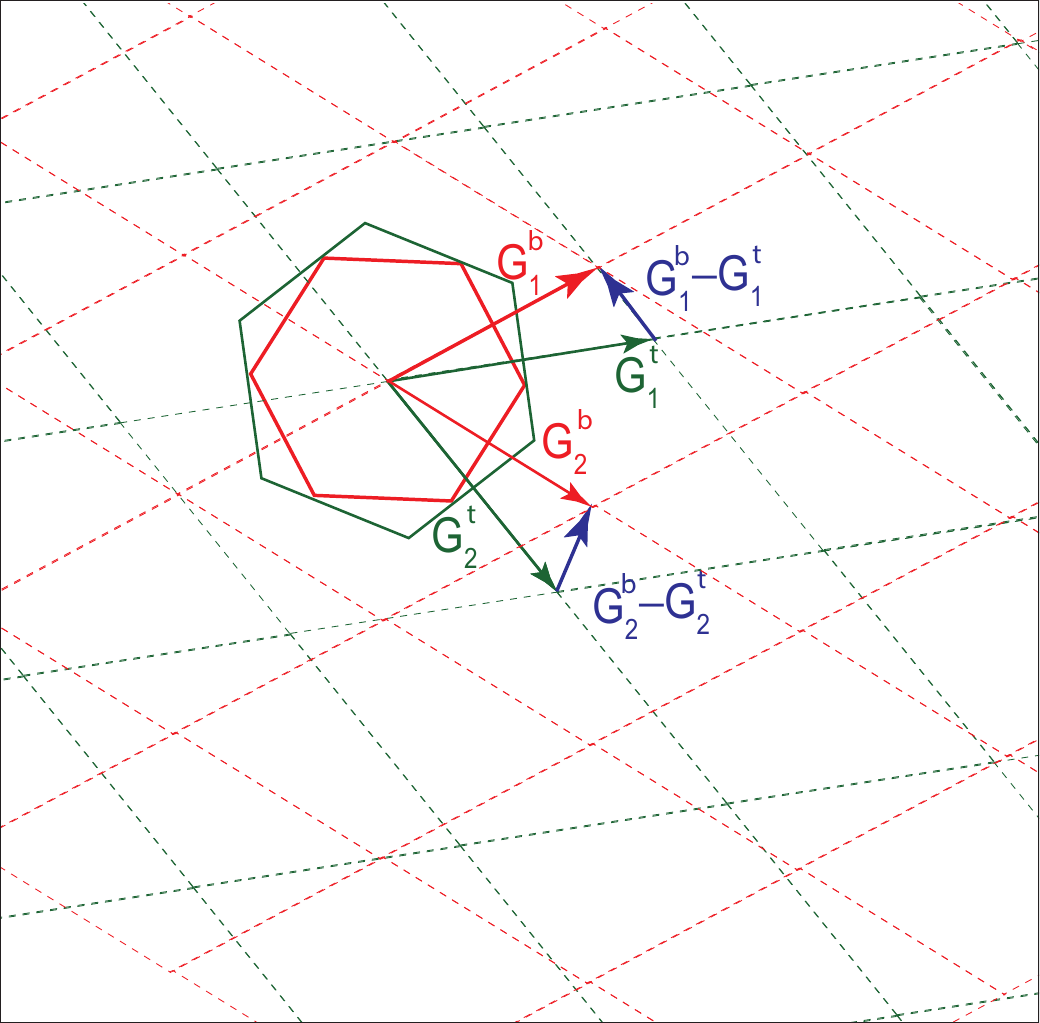}
			\small{\caption{  A schematic showing the supermoir\'{e} reciprocal lattice vectors $\vec{G_1^b}-\vec{G_1^t}$ and $\vec{G_2^b}-\vec{G_2^t}$ . \label{fig:smform}}}
		\end{center}
	\end{figure}

	Using Eqn.~\ref{Eqn:nsm}, we find that for $\theta{_b}$ and $\theta{_t}$ in opposite direction, the supermoir\'{e} gap should open at a number density $0.5 \times 10^{16}~\mathrm{m^{-2}}$. On the other hand, for $\theta{_b}$ and $\theta{_t}$ in the same relative direction, the supermoir\'{e} gap should open at a number density $0.38 \times 10^{16}~\mathrm{m^{-2}}$. The second value matches very well with our experimentally identified gap location of $n_s = 0.39 \times 10^{16}~\mathrm{m^{-2}}$. We thus conclude that the hBN at the top and the bottom have the same relative twist angles with the intervening BLG. We summarize the observations of this section in the following table:

	\begin{table}[h!]
		\begin{center}
			\begin{tabular}{|l|c|c|c|} \hline
				Interface & number density $(n)$ & moir\'{e} wavelength $(\lambda)$    & relative twist angle $(\theta)$ \\ \hline
				Bottom hBN and BLG & $n_b=2.36\times 10^{16}~\mathrm{m^{-2}}$ & $\lambda{_b} = 13.97~\mathrm{nm}$ & $\theta{_b}$ = $0.03^\circ$\\ \hline
				Top hBN and BLG & $n_{t} = 2.80\times 10^{16}~\mathrm{m^{-2}}$& $\lambda{_t} = 12.84~\mathrm{nm}$  &   $\theta{_t}$ = $0.44^\circ$\\ \hline
			\end{tabular}
			\label{table:angles}
			\caption{Calculated moir\'{e} wavelengths, number density at SDP, and moir\'{e} angles at the two interfaces of the BLG with hBN.  }
		\end{center}
	\end{table}

	\section{S3.~\hspace{0.2cm}Continuum Hamiltonian}
	\subsection{A.~\hspace{0.2cm}Model Description}

	The higher-order fractal gaps observed in the experiments can be theoretically understood by calculating the energy spectrum using the effective continuum model for hBN/BLG/hBN system with the twist angles, $\theta_{t}$ and $\theta_{b}$ for the top and bottom hBN/graphene respectively. In general, generic pairs of $\theta_{b}$ and $\theta_{t}$ leads to two incommensurate moir\'{e} periods. We construct the commensurate approximant close to the experimentally observed twist angles and write the effective Hamiltonian. For this quadrilayer system, we write the 8$\times$8 Bistritzer-MacDonald Continuum Hamiltonian in the  sublattice basis \{$A^{b},B^{b},A_{1},B_{1},A_{2},B_{2},A^{t},B^{t}$\} as
	\begin{eqnarray}
		H =
		\begin{bmatrix}
			H_{hBN} & U^{b\dagger}  & 0 & 0 \\
			U^{b} & H_{G} & U_{BLG}^{\dagger} & 0 \\
			0 & U_{BLG} & H_{G} & U^{t\dagger} \\
			0 & 0 & U^{t} & H_{hBN}
		\end{bmatrix}
	\end{eqnarray}

	where $A^{b}, B^{b}, A_{1}, B_{1}, A_{2}, B_{2}, A^{t}, B^{t}$ are sublattice sites of bottom hBN, two graphene layers, and top hBN respectively. The BLG is AB stacked such that $A_1$ and $B_2$ are vertically aligned. $H_{G}$ and $H_{hBN}$ are the Hamiltonian of graphene and hBN. The off-diagonal blocks, $U^{b}$, $U^{t}$ and $U_{BLG}$ are the interlayer potentials for top hBN/graphene, bottom hBN/graphene and bilayer graphene, respectively.

	We reduce the 8$\times$8 Hamiltonian to 4$\times$4 effective Hamiltonian, eliminating the sublattice basis of hBN using second-order perturbation and writing it as:
	\begin{eqnarray}
		H_{eff} =
		\begin{bmatrix}
			H_{G}+V_{hBN}^{b} & U_{BLG}^{\dagger} \\
			U_{BLG}                & H_{G}+V_{hBN}^{t}
		\end{bmatrix}
	\end{eqnarray}
	where in the low-energy limit, $V_{hBN}^{\ell} = U^{\ell\dagger}(-H_{hBN})^{-1}U^{\ell}
	= v_{0} + v_{1}e^{\iota\xi\vec{G}_{1}^{\ell}.\vec{r}}+
	v_{2}e^{\iota\xi\vec{G}_{2}^{\ell}.\vec{r}}+v_{3}e^{\iota\xi\vec{G}_{3}^{\ell}.\vec{r}}$ with $\ell= b,t$. $\xi$  is the valley index. $\vec{G}_{1}^{\ell}$ and $\vec{G}_{2}^{\ell}$ are reciprocal lattice basis vectors of the $\ell$ moir\'{e} and $\vec{G}_{3}^{\ell} = -\vec{G}_{1}^{\ell}-\vec{G}_{2}^{\ell}$. The twist angle dependence is reflected in the $\vec{G}_{1,2}^{\ell}$ in the potential.

	The $v_{0}$, $v_{1}$, $v_{2}$ and $v_{4}$ are
	$V_{0}\begin{bmatrix}
		1 & 0 \\
		0 & 1
	\end{bmatrix}$, $V_{1}e^{\iota\xi\phi}\begin{bmatrix}
		1 & \omega^{-\xi} \\
		1 & \omega^{-\xi}
	\end{bmatrix}$, $ V_{1}e^{\iota\xi\phi}\begin{bmatrix}
		1 & \omega^{\xi} \\
		\omega^{\xi} & \omega^{-\xi}
	\end{bmatrix}$ and
	$V_{1}e^{\iota\xi\phi}\begin{bmatrix}
		1 & 1 \\
		\omega^{-\xi} & \omega^{-\xi}
	\end{bmatrix}$ respectively, with $\omega = e^{\iota2\pi/3}$, $V_{0}=29$~meV, $V_{1}=21$~meV and $\phi=-0.29$~rad.

	\subsection{B.~\hspace{0.2cm} Band Structure Calculation}
	\begin{figure}[t]
		\begin{center}
			\includegraphics[width=0.9\columnwidth]{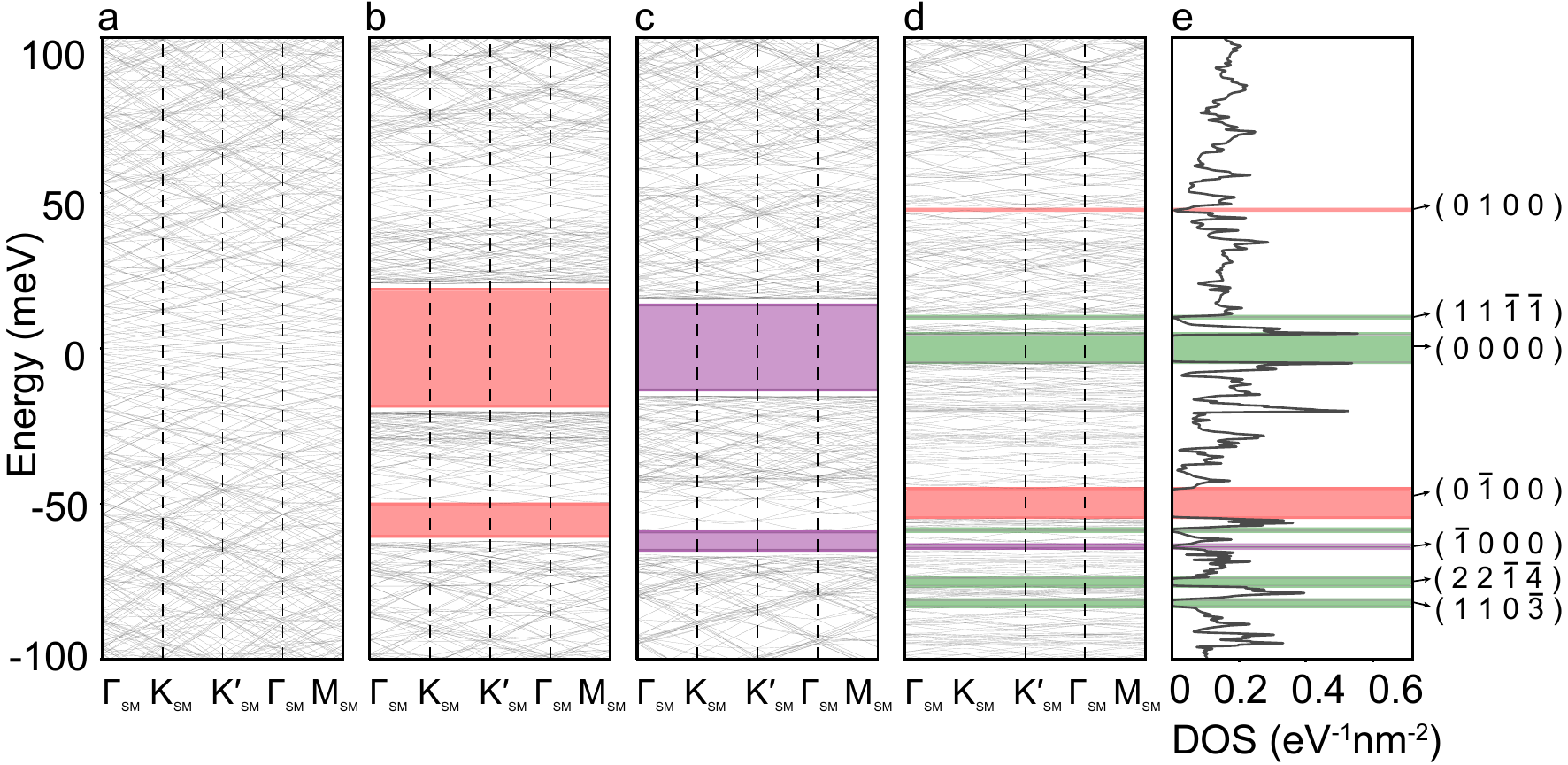}		\small{\caption{\textbf{Band Structure and density of states for the system with $\theta_{b}=0.026^{\circ}$ and $\theta_{t}=0.44^{\circ}$.} (Plot of the band structure of bilayer graphene with (a) no moir\'{e} potential, (b) top moir\'{e} potential, (c) bottom moir\'{e} potential, and (d) double moir\'{e} potential along the path $\mathrm{\Gamma_{SM}-K_{SM}-K'_{SM}-\Gamma_{SM}-M_{SM}}$ of the supermoir\'{e} BZ. (e) Plot of the DOS versus energy calculated for the supermoir\'{e} band structure in (d). The primary first-order gaps are shown in red and purple in (b), (c), (d), and (e). The secondary higher-order gaps are marked in green in (d) and (e).}
				\label{fig:Bands}}
		\end{center}
	\end{figure}
	We write the Hamiltonian in the  bilayer graphene basis $\{|\vec{{q}_{m_{1},m_{2}}},A_{1}\rangle, |\vec{{q}_{m_{1},m_{2}}},B_{1}\rangle,|\vec{{q}_{m_{1},m_{2}}},A_{2}\rangle,$
	\\ $ |\vec{{q}_{m_{1},m_{2}}},B_{2}\rangle\}$ where $\vec{q}_{m_{1},m_{2}} = \vec{k}+m_{1}\vec{G}_{1}^{SM}+m_{2}\vec{G}_{2}^{SM}$. $\vec{k}$ is the vector in first supermoir\'{e} Brillouin zone and $\vec{G}_{1}^{SM}$ and $\vec{G}_{2}^{SM}$ are the basis vectors of the supermoir\'{e} reciprocal lattice. A spherical $k$-space cutoff is chosen such that $|\vec{{q}_{m_{1},m_{2}}}| \leq \vec{q}_{c}$. We take the cutoff, $\vec{q}_{c}=2|\vec{G}_{1}^{t}|$. This choice for the cutoff ensures the convergence of the band structure. The band structure is obtained by diagonalizing the continuum Hamiltonian across the path $\Gamma_{SM}-K_{SM}-K'_{SM}-\Gamma_{SM}-M_{SM}$ in the first supermoir\'{e} Brillouin zone. The band structures of bare bilayer graphene, bilayer graphene forming top moir\'{e} with hBN, bilayer graphene forming bottom moir\'{e} with hBN, and hBN/BLG/hBN double moir\'{e} are shown in Fig.~\ref{fig:Bands}(a-d).

	\subsection{C.~\hspace{0.2cm} Density of states (DOS) versus number density}
\begin{figure*}[t]
		\includegraphics[width=0.85\columnwidth]{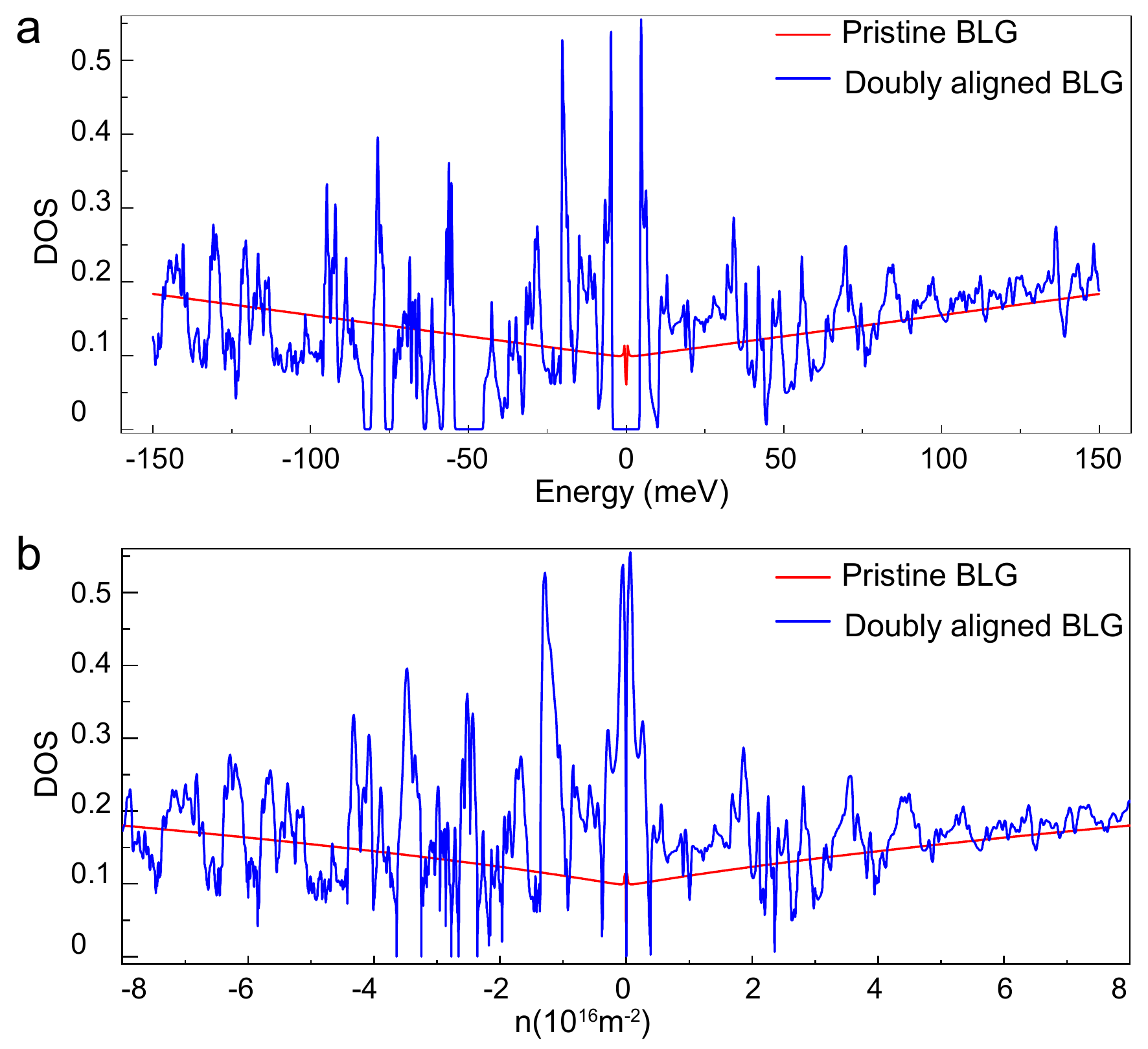}		\small{\caption{\textbf{Comparison of DOS.} Plots comparing the calculated DOS versus (b) energy and (b) versus electron density for the pristine BLG (red curve) and doubly aligned BLG with hBN with $\theta_{b}=0.026^{\circ}$ and $\theta_{t}=0.44^{\circ}$ (blue curve). }
			\label{fig:blgdos}}
	\end{figure*}

	The density of states, $g(E)$ is calculated as
	\begin{eqnarray}
		g(E) = \frac{4}{S}\sum_{n\vec{k}}w_{n\vec{k}}\delta(E-E_{n\vec{k}})
		\label{eqn:s8}
	\end{eqnarray}
	where $S$ is the real space area of supermoir\'{e} first BZ. We used a uniform sampling of the supermoir\'e BZ of 200$\times$200 $k$-points to calculate the DOS. The factor of 4 in Eq.~\ref{eqn:s8} is to account for the four-fold valley and spin degeneracy. The delta function, $\delta(E-E_{n\vec{k}})$, is approximated as a gaussian with broadening 0.17~meV corresponding to the measurement temperature of 2~K. The DOS as a function of $E$ (with the gaps highlighted) is shown in Fig.~\ref{fig:Bands}(e). We also overlay the DOS of the pristine bilayer graphene on the DOS of doubly aligned BLG with hBN (Fig.~\ref{fig:blgdos}). The DOS of pristine BLG is linear at finite $E$. The moir\'{e} potential results in multiple dips and gaps over the linear DOS of pristine BLG. To compare with experimental results,  we convert the calculated DOS to be a function of the number density $n$ using $n(E) = 4\sum_{n\vec{k}}w_{n\vec{k}}\Theta(E-E_{n\vec{k}})$, where $\Theta(E-E_{n\vec{k}})$ is the Heaviside step function.

	\section{S4.~\hspace{0.2cm}Determining the zone quantum numbers for the gaps}
\begin{figure}[t]
		\begin{center}
			\includegraphics[width=\columnwidth]{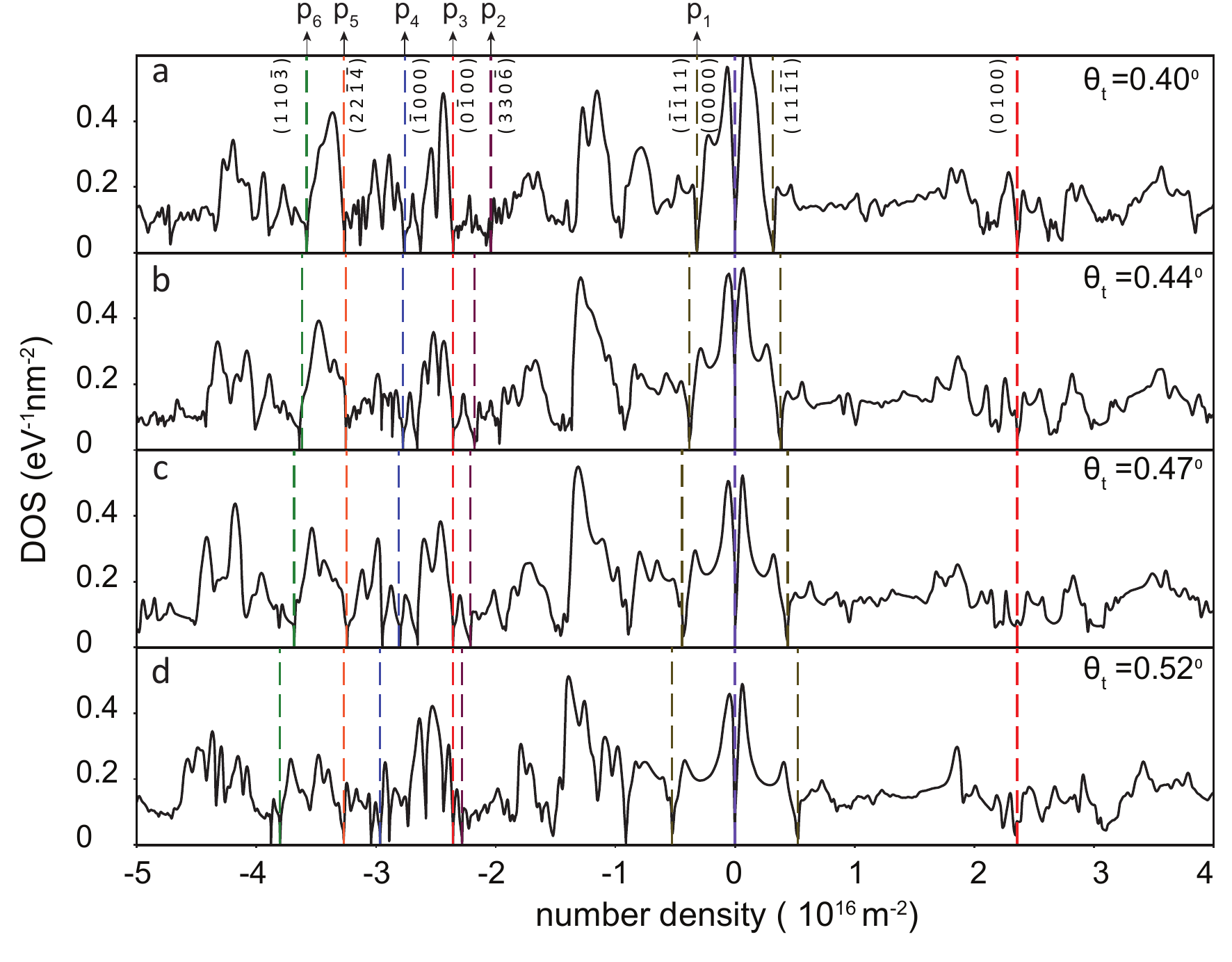}
			\small{\caption{\textbf{DOS versus n for 4 different angles}. (a)-(d) shows the DOS versus n for $\theta_{t} = 0.40^{\circ}, \  0.44^{\circ}, \ 0.47^{\circ}, \ 0.52^{\circ}$ respectively. The positions of densities corresponding to the gaps, with quantum numbers marked, are shown with dotted lines.
				}
				\label{fig:quantumno}}
		\end{center}
	\end{figure}

	The number density corresponding to a gap is related to the areas $A_{1},A_{2},A_{3},A_{4}$ by the set of integers $m_{1},m_{2},m_{3},m_{4}$ as:
	\begin{eqnarray}
		n = 4(m_{1}A_{1}+m_{2}A_{2}+m_{3}A_{3}+m_{4}A_{4})/(2\pi)^{2}
		\label{supp:numberdensity_area}
	\end{eqnarray}
	where areas are defined as the independent subset of the cross products of moir\'{e} reciprocal lattice vectors. There are, thus, four unknowns in the equation. Also, the electron number density corresponding to band gaps evolves continuously with changes in the twist angle (Fig.~\ref{fig:quantumno}). We consider four different twist angles ($\theta_{t} = 0.40^{\circ}, 0.44^{\circ},0.47^{\circ},0.52^{\circ}$) to find the integers for the gaps observed in $\theta_{t} = 0.44^{\circ}$. For a commensurate system, $A_{1}, A_{2}, A_{3}, \textrm{and} A_{4}$ have the highest common factor $A_{SM} = \left|\vec{G}_{1}^{SM} \times \vec{G}_{2}^{SM}\right|$. As a result, they can be written as an integral multiple of $A_{SM}$ via the relation $A_{i} = s_{i}A_{SM}$ with integers $s_{i}$ (i = 1, 2, 3, 4). We also characterize $n_{e}$ for a gap by an integer $p$, such that $n_{e} = pA_{SM}/(2\pi)^{2}$. $p$ equals the number of bands from the CNP to the gap with density $n_e$.

	Substituting the areas and number densities in terms of an integral multiple of supermoir\'{e} areas in the Eq. (\ref{supp:numberdensity_area}), we get the Diophantine equation~\cite{PhysRevB.104.035306}:
	\begin{equation}
		p = m_{1}s_{1} + m_{2}s_{2} + m_{3}s_{3}+ m_{4}s_{4}
	\end{equation}
	We construct four Diophantine equations for each number density (using the four twist angles mentioned above). The system of equations is written as $SX=P$, where S is 4$\times$4 matrix shown as the columns $s_1$ to $s_4$ in the \ref{table:braggeqns}. P is the column $p_1$ to $p_6$ for six gaps, respectively. The solution of this set of equations is unique and gives zone quantum numbers, $X = (m_{1}, m_{2}, m_{3}, m_{4})$ for each gap.

	\begin{table}[b]
		\centering
		\begin{tabular}{@{}|c||c|c|c|c||c|c|c|c|c|c||c||} \hline
			\hspace{0.3cm} $\theta_{t}$ \hspace{0.3cm} & \hspace{0.3cm} $s_{1}$ \hspace{0.3cm} & \hspace{0.3cm} $s_{2}$ \hspace{0.3cm} & \hspace{0.3cm} $s_{3}$\hspace{0.3cm} & \hspace{0.3cm} $s_{4}$ \hspace{0.3cm} & \hspace{0.3cm} $p_{1}$ \hspace{0.3cm} & \hspace{0.3cm} $p_{2}$\hspace{0.3cm} & \hspace{0.3cm} $p_{3}$ \hspace{0.3cm} & \hspace{0.3cm} $p_{4}$ \hspace{0.3cm} & \hspace{0.3cm} $p_{5}$ \hspace{0.3cm} & \hspace{0.3cm} $p_{6}$ \hspace{0.3cm} & $A_{SM} (\times10^{-2}nm^{-2})$ \\ \hline
			0.40 & 61 & 52 & 42 & 64 & -7 & -45 & -52 & -61 & -72 & -79 & 0.441 \\
			0.44 & 93 & 79 & 61 & 98 & -13 & -72 & -79 & -93 & -109 & -122 & 0.297\\
			0.47 & 19 & 16 & 12 & 20 & -3 & -15 & -16 & -19 & -22 & -25 & 1.484\\
			0.52 & 39 & 31 & 23 & 40 & -7 & -30 & -31 & -39 & -43 & -50 & 0.7491\\
			\hline

		\end{tabular}
		\label{table:braggeqns}
		\caption{Tabulating the Diophantine equations for the gaps of 4 different angles ($\theta_{t}$) as the augmented matrix form of the system of equations, (S|P), where S is the $4\times4$ matrix from column ($2-5$) and the P in column ($6-11$) are the integer equivalents of the number densities for 6 gaps. The last column ($A_{SM}$) is the reciprocal space area of supermoir\'e}
	\end{table}

	\section{S5.~\hspace{0.2cm}Quasi Brilluoin Zones}
	\begin{figure}[H]
		\includegraphics[width=0.9\columnwidth]{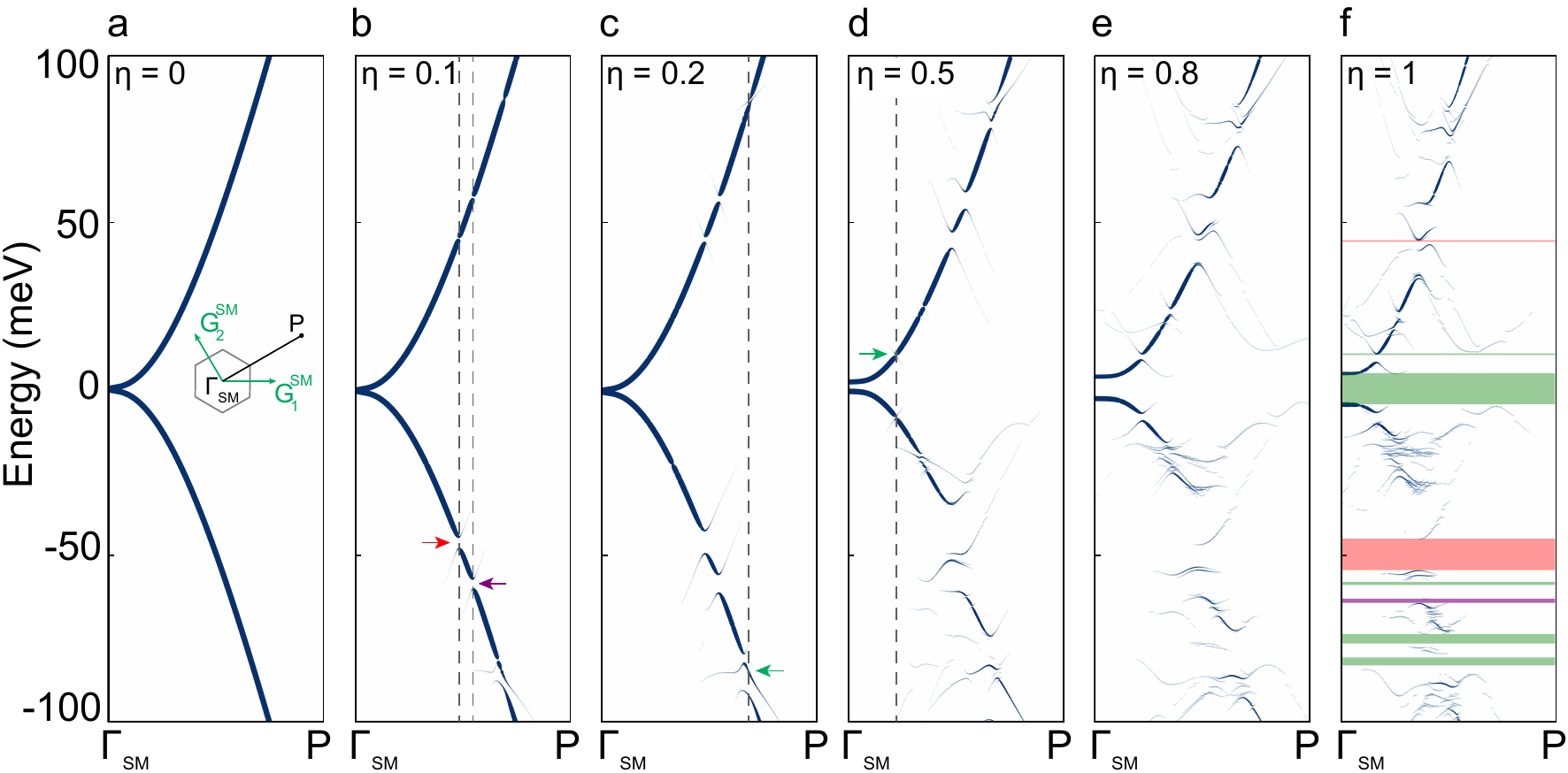}
		\small{\caption{\textbf{Band structure for a range of moir\'{e} potential strengths parameterised by $\eta$.} Band structure across the path $\mathrm{\Gamma_{SM}}$ to P unfolded to the unit cell of bilayer graphene. (a) The parabolic-like bands of bilayer graphene with no moir\'{e} potential. (b), (c), (d), and (e) show the band structure with $\eta$ varying from 0.2 to 1.0. The calculated spectral weights are shown as purple dots. The gap opening points can be traced back at different $\eta$. Vertical dashed lines represent the k point at which the gap opens along this path in the weak potential limit. }
			\label{fig:unfold}}
	\end{figure}
	Below we describe the procedure to construct the qBZ by modulating the moir\'{e} potential strength in Hamiltonian by a factor of $\eta$:
	\begin{eqnarray}
		H_{eff} =
		\begin{bmatrix}
			H_{G}+\eta V_{hBN}^{b} & U_{BLG}^{\dagger} \\
			U_{BLG}                & H_{G}+\eta V_{hBN}^{t}
		\end{bmatrix}
	\end{eqnarray}
	We plot the unfolded the band structure across the path $\mathrm{\Gamma_{SM}}$ to a point P in $k_x-k_y$ plane for $\eta = 0.0, 0.1, 0.2, 0.5, 0.8$ \& $1.0$ (Fig.~\ref{fig:unfold}). At $\eta=0$, we see the bilayer graphene parabolic dispersion in the unfolded band structure, and for $\eta=0.1,0.2,0.5,0.8,1.0$, we see gaps appearing in the dispersion with the increasing $\eta$. We trace the gap-opening points along this path. We repeat the procedure to mark the gap opening points on the whole $k_x-k_y$ plane. The qBZs thus constructed are shown in Fig.(3) of the main text.

	\section{S6.~\hspace{0.2cm} Comparison of qBZ in SLG and BLG}

	\begin{figure}
		\begin{center}
			\includegraphics[width=0.45\textwidth]{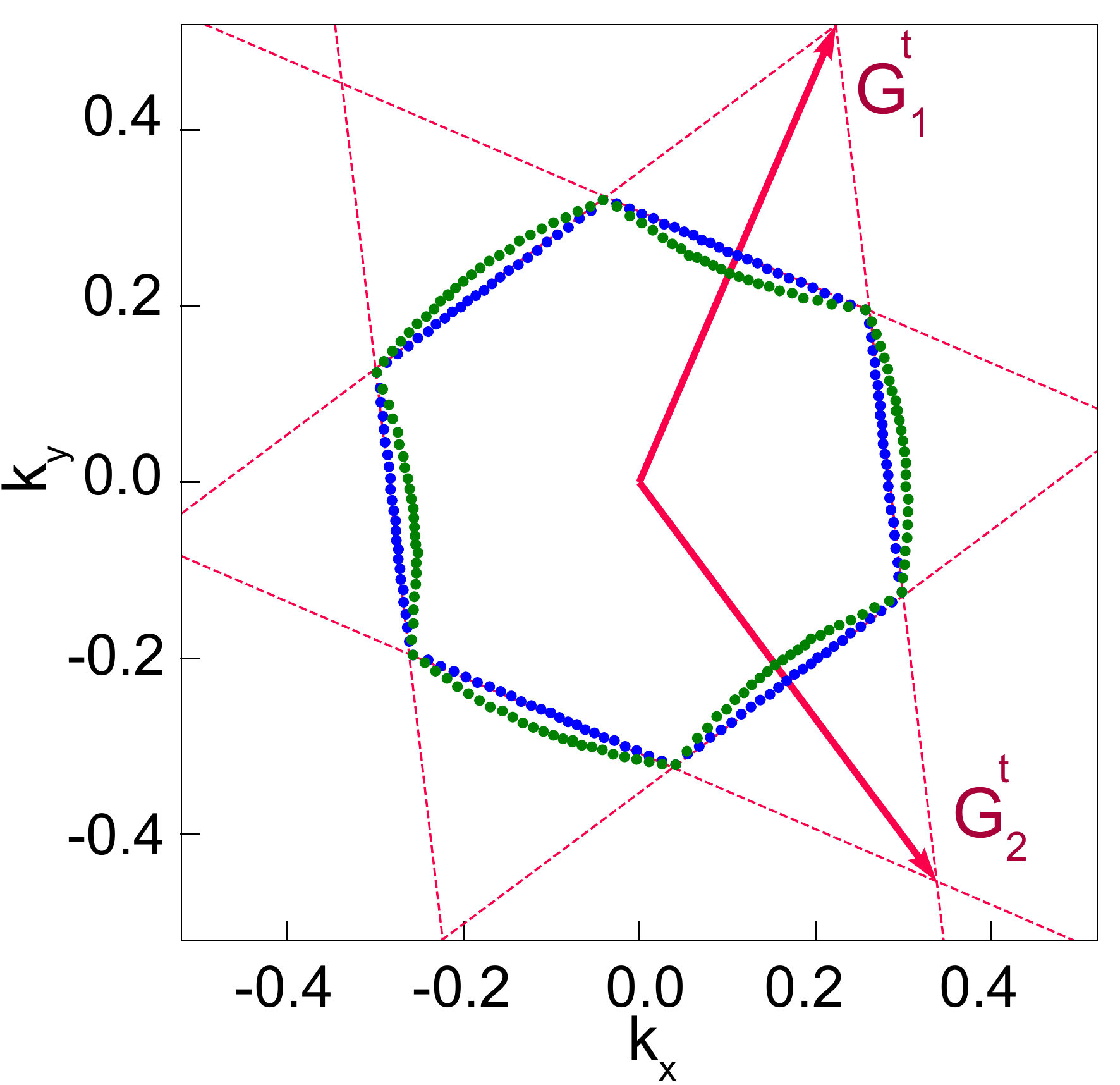}
			\small{\caption{ \textbf{BZs for SLG and BLG aligned with hBN}. Brillouin Zones constructed by unfolding the primary moir\'{e} gaps for twist angle of $\theta_{t}=0.44^{\circ}$. The blue (green) filled circles show the BZ of single-layer graphene/hBN (bilayer graphene/hBN) moir\'{e}. The area of each zone is equal to the corresponding number density at which the gap is observed. }
				\label{fig:BZconduction}}
		\end{center}
	\end{figure}

\begin{figure}[H]
		\begin{center}
			\includegraphics[width=1\textwidth]{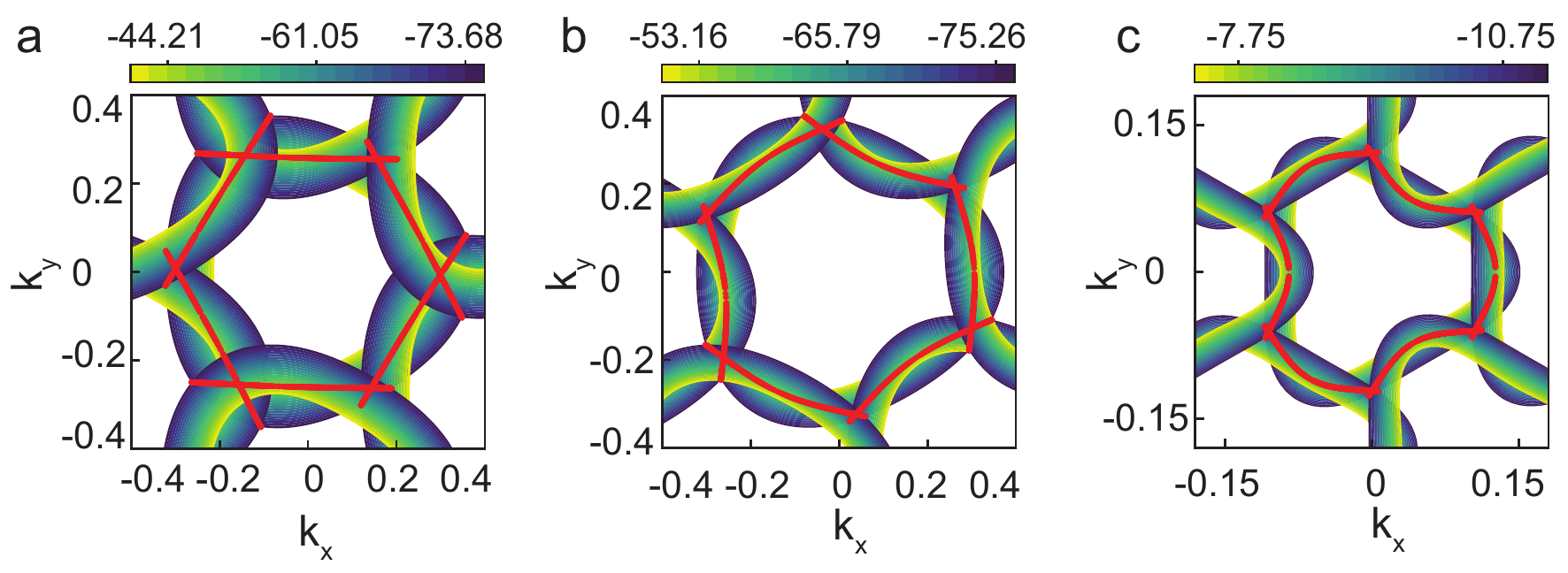}
			\small{\caption{\textbf{BZ construction using intersection of iso-energy contours.} (a), (b), and (c) show the BZ constructed by the intersection of iso-energy contours placed at $\mathrm{\vec{G_{i}^{b}},\vec{G_{i}^{t}},\vec{G_{i}^{b}}-\vec{G_{i}^{t}}}$ with the dispersion centred at (0,0) in $\mathrm{k_{x}-k_{y}}$ plane. Here $i=1$ to $6$. $\mathrm{k_{x}}$ and $\mathrm{k_{x}}$ are in nm$^{-1}$ and colorbars in meV. These zones agree with those constructed via the unfolding procedure in Fig.(3) of the main manuscript.}
				\label{fig:contour}}
		\end{center}
	\end{figure}

	Some of the qBZs constructed for the bilayer graphene doubly aligned with hBN are distorted hexagons (see Fig.3, main text). To understand their origin, we compare the qBZ for single moir\'{e} formed between single-layer graphene (SLG) and hBN with that formed between bilayer graphene and hBN. In both cases, the angular misalignment between the graphene and the hBN was set to be $\theta_{t} = 0.44^{\circ}$. The resultant qBZ are plotted in Fig.~\ref{fig:BZconduction}. The qBZ in the case of SLG is perfectly hexagonal; this can be understood by considering the circular iso-energy contours in graphene dispersion. We translate the contours by vectors, $\vec{G}_{i}^{t}$ ($i=1$ to $6$). The intersection of the iso-energy contours gives the hexagon. For the BLG, on the contrary, one has triangular symmetric constant energy contours due to the trigonal warping term in the Hamiltonian. A similar intersection procedure results in distorted hexagons. (Fig.~\ref{fig:contour}).

	\section{S7.~\hspace{0.2cm} Analysis of previously published data on SLG supermoir\'e}
	\begin{figure}
		\begin{center}
			\includegraphics[width=1\textwidth]{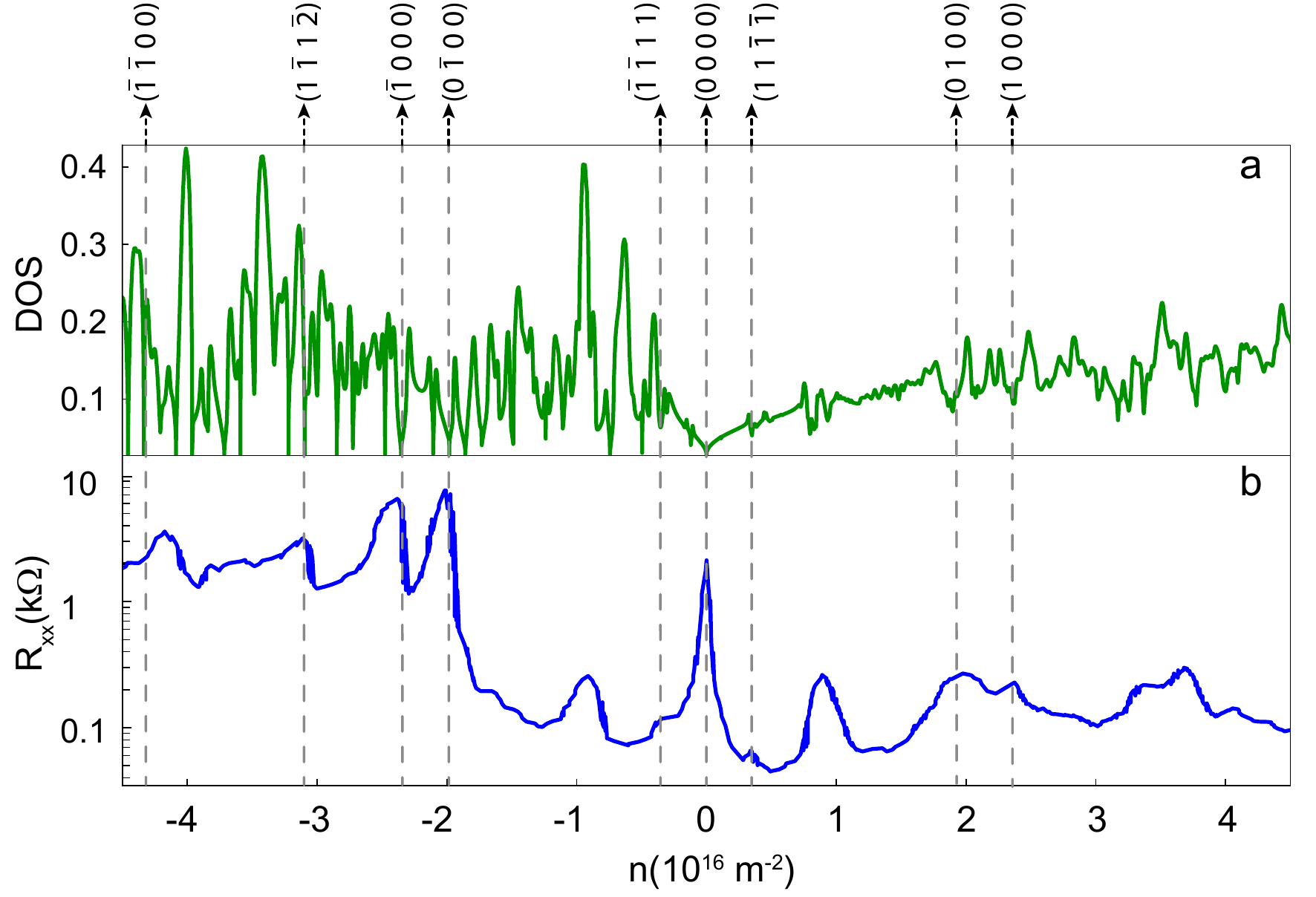}
			\small{\caption{\textbf{Experimentally measured and theoretically calculated Bragg gaps} (a) Plot of the density of states (DOS) versus $n$ for twist angle of $\theta_{b}=0^{\circ}$ and $\theta_{t}=0.407^{\circ}$. (b) Plot of Longitudinal resistance R$_{xx}$ versus $n$ from Ref.~\cite{doi:10.1126/sciadv.aay8897}. The vertical dashed lines mark the major dips in the DOS and corresponding peaks in the resistance curve.}
				\label{fig:Singlelayer}}
		\end{center}
	\end{figure}
	In a system of graphene aligned with a single hBN, a single moir\'{e} periodicity is typically generated, resulting in one secondary Dirac point on both the electron and hole side. However, when graphene is doubly aligned with hBN, it produces multiple gaps in addition to the primary gaps of the top and bottom moir\'{e}. This is due to the interference between the two moir\'{e} potentials. These gaps appear as peaks in the resistance versus carrier density curve during electronic transport measurements, as demonstrated in bilayer graphene data (as shown in the main text Fig.(1)) and previous studies from other groups in single layer graphene(SLG)~\cite{doi:10.1126/sciadv.aay8897}.

	The data published in Ref.\cite{doi:10.1126/sciadv.aay8897} are reproduced  in Fig.\ref{fig:Singlelayer}. This pioneering study used a first-order approach to explain the origin of multiple peaks in the system. However, some prominent peaks, for example, those at $\mathrm{n\approx\pm 3.10\times 10^{16}~\mathrm{m^{-2}}}$ and $\mathrm{n\approx\pm 4.35\times 10^{16}~\mathrm{m^{-2}}}$,  remained unexplained.
	We show that the origin of these peaks can be understood under a generalized continuum model approach. The effective continuum Hamiltonian for hBN/SLG/hBN is:
	\begin{eqnarray}
		H = H_{G} + V_{hBN}^{b}+V_{hBN}^{t}
	\end{eqnarray}
	We use the lattice constants for graphene and hBN as 0.2464 and 0.2504 nm, respectively, to have the lattice mismatch of 1.6$\%$ (as mentioned by the authors of Ref.\cite{doi:10.1126/sciadv.aay8897}).
	Fig.~\ref{fig:Singlelayer} shows the theoretical density of states for the given twist angles $\theta_{b}=0^{\circ}$ and $\theta_{t}=0.407^{\circ}$.The areas $A_i$ for the experimentally obtained twist angle are $\mathrm{0.233, \ 0.196, \ 0.147}$ and $\mathrm{0.245}$ nm$^{-2}$. The gaps at $n= -1.98\times 10^{16}~\mathrm{m^{-2}}$ and $n= -2.36\times 10^{16}~\mathrm{m^{-2}}$ corresponds to quantum number $(0,\bar1,0,0)$ and $(\bar1,0,0,0)$ respectively. The peak in $R_{xx}$ at at $n= -0.37\times 10^{16}~\mathrm{m^{-2}}$ corresponds to quantum number $(\bar1,\bar1,1,1)$ and is identified to be the supermoir\'{e} gap. Peaks that went unexplained in the original publication at $n= -3.10\times 10^{16}~\mathrm{m^{-2}}$ and $n= -4.35\times 10^{16}~\mathrm{m^{-2}}$ can now be understood to arise due to the noticeable zeros in the calculated density of states and correspond to the quantum number $(1,\bar1, 1,\bar2)$  and $(\bar1,\bar1,0,0)$ respectively.
	\begin{table}
		\centering
		\begin{tabular}{@{}|c||c|c|c|c||c|c|c|c|c|c||c||} \hline
			\hspace{0.3cm} $\theta_{t}$ \hspace{0.3cm} & \hspace{0.3cm} $s_{1}$ \hspace{0.3cm} & \hspace{0.3cm} $s_{2}$ \hspace{0.3cm} & \hspace{0.3cm} $s_{3}$\hspace{0.3cm} & \hspace{0.3cm} $s_{4}$ \hspace{0.3cm} & \hspace{0.3cm} $p_{1}$ \hspace{0.3cm} & \hspace{0.3cm} $p_{2}$\hspace{0.3cm} & \hspace{0.3cm} $p_{3}$ \hspace{0.3cm} & \hspace{0.3cm} $p_{4}$ \hspace{0.3cm} & \hspace{0.3cm} $p_{5}$ \hspace{0.3cm} & \hspace{0.3cm} $A_{SM} (\times10^{-2}nm^{-2})$ \\ \hline
			0.33 & 28 & 25 & 20 & 30 & -3 & -25 & -28 & -37 & -53 & 0.784 \\
			0.36 & 31 & 27 & 21 & 33 & --4 & -27 & -31 & -41 & -58 & 0.726\\
			0.41 & 19 & 16 & 12 & 20 & -3 & -16 & -19 & -25 & -35 & 1.225\\
			0.47 & 61 & 49 & 35 & 63 & -12 & -49 & -61 & -79 & -110
			& 0.4\\
			0.54 & 4 & 3 & 2 & 4 & -1 & -3 & -4 & -5 & -7 & 6.542 \\

			\hline

		\end{tabular}
		\label{table:braggeqns}
		\caption{Tabulating the Diophantine equations for the gaps of 5 different angles ($\theta_{t}$) as the augmented matrix form of the system of equations, (S|P), where S is the $5\times4$ matrix from column ($2-5$) and the P in column ($6-11$) are the integer equivalents of the number densities for 5 gaps with quantum numbers $(\bar1,\bar1,1,1),(0,\bar1,0,0),(\bar1,0,0,0),(1,\bar1, 1,\bar2)$  and $(\bar1,\bar1,0,0)$ respectively. The last column ($A_{SM}$) is the area of the reciprocal space  supermoir\'e cell.}
	\end{table}

	\clearpage
	\bibliography{Arxiv}

\end{document}